\documentclass[3p,preprint,final]{elsarticle}
\usepackage{ecrc}

\volume{00}

\firstpage{1}

\runauth{M. Rigaki and S. Garcia}

\jid{cose}

\usepackage[cmex10]{amsmath}
\usepackage{bbm}
\usepackage{tabularx}
\usepackage{graphicx}
\usepackage{subcaption}
\usepackage{booktabs}
\usepackage{hyperref}
\usepackage{amssymb}
\usepackage{adjustbox}
\usepackage{multirow}

\hypersetup{
colorlinks=true,
linkcolor=red,
citecolor=green,
filecolor=magenta,
urlcolor=blue
}

\begin{document}
\begin{frontmatter}


\title{Stealing and Evading Malware Classifiers and Antivirus at Low False Positive Conditions}


\author[ctu]{M. Rigaki\corref{cor1}}
\ead{maria.rigaki@fel.cvut.cz}

\author[ctu]{S. Garcia}
\ead{sebastian.garcia@fel.cvut.cz}
\address[ctu]{Czech Technical University in Prague, Department of Computer Science, Prague, Czech Republic}

\cortext[cor1]{Corresponding author}

\begin{abstract}
Model stealing attacks have been successfully used in many machine learning domains, but there is little understanding of how these attacks work against models that perform malware detection. Malware detection and, in general, security domains have unique conditions. In particular, there are very strong requirements for low false positive rates (FPR). Antivirus products (AVs) that use machine learning are very complex systems to steal, malware binaries continually change, and the whole environment is adversarial by nature. 

This study evaluates active learning model stealing attacks against publicly available stand-alone machine learning malware classifiers and also against antivirus products. The study proposes a new neural network architecture for surrogate models (dualFFNN) and a new model stealing attack that combines transfer and active learning for surrogate creation (FFNN-TL). We achieved good surrogates of the stand-alone classifiers with up to 99\% agreement with the target models, using less than 4\% of the original training dataset. Good surrogates of AV systems were also trained with up to 99\% agreement and less than 4,000 queries. The study uses the best surrogates to generate adversarial malware to evade the target models, both stand-alone and AVs (with and without an internet connection). Results show that surrogate models can generate adversarial malware that evades the targets but with a lower success rate than directly using the target models to generate adversarial malware. Using surrogates, however, is still a good option since using the AVs for malware generation is highly time-consuming and easily detected when the AVs are connected to the internet.

\end{abstract}


\end{frontmatter}

\section{Introduction}
\label{sec:introduction}
In recent years there has been a strong interest in model stealing attacks to understand how models can be extracted and how model information can be leaked. These attacks are relevant to anyone deploying machine learning models since they are designed to either steal the intellectual property of the model owner or to use the stolen model to compromise the privacy and integrity of the target model. Model stealing attacks were successful to some extent, implying that machine learning models and their deployment must be adapted and consider new risks. However, there needs to be more extensive analysis and understanding of model stealing for malware detection tasks and how it can be used later as a stepping stone for other attacks.

Despite the importance of correctly detecting malware, there has yet to be research exploring the best way to extract detection models. The task of stealing malware classifiers has one additional constraint compared to attacks in traditional AI domains: it requires that both the targets and the surrogates perform well in terms of detection while exhibiting very low false positives. Low false positives are required to avoid blocking or removing files that may be important for the system's use. Another reason is to reduce user fatigue from false alarms. In addition, the security domain has other characteristics, such as that antivirus being very complex black-box systems with many parts and detection techniques.

Exploring and understanding model stealing attacks is also significant because they can be used in other downstream attacks, such as evading detection models by creating adversarial malware. Detection evasion is not a new topic; dozens of different techniques exist to modify malware. However, creating adversarial malware with surrogate models can significantly increase attackers' success in evading detection. 

Furthermore, antivirus systems are highly complex systems that use multiple components. In the industry and in academic research, they are considered black boxes, and companies do not tend to share detailed information about their internals. Reverse engineering each product can be very time-consuming, but having access to a surrogate that models their behavior can potentially be an exciting avenue for attackers. 

This research has two stages: first, to steal stand-alone ML models and antivirus (AVs) detection models by creating good surrogates on low FPR conditions; second, to use those surrogates to create adversarial malware and evaluate their evasion performance. 

The first stage creates surrogates by combining different active learning algorithms and sampling strategies. The four sampling strategies used for the stealing attacks are \textit{random sampling}, \textit{entropy}, \textit{entropy+k-medoids}, and \textit{Monte Carlo Dropout+entropy}. The last two are our new variations and contributions. We use two baseline surrogates: one based on neural networks and one on lightGBM. We also propose two new neural network-based approaches for creating surrogates. The first one is a new architecture called dualFFNN, which takes advantage of the existence of true labels and uses a skip connection to add stability during training. The second is called FFNN-TL and combines data-based transfer learning with active learning. We focused on creating high-performance surrogates based on neural networks because they provide more flexibility than other high-performing models, such as gradient boosting trees. Neural network models are easier to update incrementally instead of training with all the data from scratch, and they can be partially used in other tasks such as transfer learning. 
The created surrogates are compared by their performance agreement with the target models and their accuracy, measured at fixed FPR levels of 0.01 or lower. 

Regarding the target models to steal, we work with two types. The first type is stand-alone machine learning models publicly available on the internet; one released as part of the Ember 2018 dataset~\cite{anderson2018ember}, and two released as part of the Sorel20m dataset~\cite{harang2020sorel20m}. The second type of models to steal are four of the top 10 AVs in the industry for home users, tested offline and online. 

Results of this first stage show that for stand-alone targets, our best dualFFNN surrogate achieves an agreement with the target from~97.8\% to~99\% and, on average, is within~1\% of the accuracy obtained by the target models (in some cases having better accuracy than the stand-alone target). For the AV targets, the FFNN-TL performs better than the dualFFNN and achieves agreement and accuracy of~97\% or more by only using 4,000 queries to the AVs or less.

The second stage of this research uses the previous models (surrogates and targets) to create adversarial malware to evade the detection of target models. This stage explores questions related to the performance of the different surrogate creation techniques. For a better comparison, the original target models are also used to create adversarial malware. Results show that the dualFFNN surrogate seems to have a good evasion capability ranging from 35\% to 70\% evasion rate against Ember and Sorel-LGB targets. In contrast, the FFNN-TL had an evasion range of 35-50\% for the same targets using both MAB and GAMMA attacks. The evasion rate of the AVs was higher when using the GAMMA attack, where most surrogates achieved more than 40\% evasion rate for AV1 and AV3. However, none of the surrogates did very well against AV2 and AV4.


This work is the first comprehensive comparison of surrogate models for adversarial malware generation to evade stand-alone machine learning models and actual antivirus products (both offline and online).

We conclude that we can create good surrogates of malware detection models at very low FPR conditions and a small number of queries to the targets using our proposed dualFFNN or FFNN-TL architectures. The impact of creating these surrogate models was explored by successfully creating adversarial malware that could evade stand-alone models and AV products. 

The main contributions of this paper are:
\begin{itemize}
    \item A new neural network architecture for surrogate creation that performs well under low FPR conditions (dualFFNN).
    \item A new surrogate creation method that combines transfer and active learning (FFNN-TL).
    \item A comparison of surrogate model creation techniques for malware detection at low FPR settings tested on both machine learning models and antivirus systems as targets.
    \item A study of the performance of the different surrogates in the downstream task of adversarial malware creation for evading machine learning models and antivirus systems. 

\end{itemize}

The rest of the paper is structured as follows: Section~\ref{sec:background-and-related-work} analyses background concepts and previous work; Section~\ref{sec:threat-model} describes the threat model; Section~\ref{sec:methodology} presents the methodology used for surrogate creation and adversarial generation; Section~\ref{sec:combining-tl-al} details the combined transfer and active learning attack;  Section~\ref{sec:surrogate_models} presents the surrogate models and our proposed dualFFNN architecture; Section~\ref{sec:models-under-attack} describes the target models; Section~\ref{sec:datasets} describes the datasets used;  Section~\ref{sec:stealing_stand_alone_ml_models} shows the experiments and results of stealing stand-alone malware detection ML models; Section~\ref{sec:stealing-av-functionality} shows the experiments and results of stealing antivirus models; Section~\ref{sec:creating-adversarial-malware} shows the results of creating adversarial malware to evade targets; Section~\ref{sec:discussion} discusses the implications of the results; and Section~\ref{sec:conclusion} presents the conclusions. The source code repository for reproducing our results can be found in~\url{https://github.com/stratosphereips/model_extraction_malware}.

\section{Background and Related Work}
\label{sec:background-and-related-work}

In model extraction or model stealing attacks, an adversary creates a new \textit{surrogate} model by smartly querying a target model and, therefore \textit{learning} from it, obtaining an equivalent performance to the stolen target model. The target model is typically a black-box or gray-box setup. The type of information taken varies from stealing the model's functionality to stealing the model's architecture, hyperparameters, optimizers, etc. 

Depending on the target model $f$, the adversary may query this model $f$ with data $X$ from what is called a \textbf{thief} dataset $\mathcal{D}_{thief}$, and retrieve classification labels or confidence vectors $\hat{y}_{target}$ from the target model. An important aspect is that these attacks are usually performed under a query budget since model queries can be costly in terms of money and time.

Following the definitions in~\cite{jagielski2020high}, in a \textbf{fidelity} attack, the adversary aims to create a surrogate model that learns the decision boundary of the target as faithfully as possible, including the errors that the target makes. These surrogate models can be used later in other tasks, such as generating adversarial samples. In a \textbf{task accuracy} attack, the adversary aims to construct a surrogate model that performs equally well \textit{or better} than the target model in a specific task such as image or malware classification. The attacker's ultimate goal affects the selection of the thief dataset, the metrics for a successful attack, and the attack strategy itself. 

One of the first model extraction attacks was proposed by Tramer et al.~\cite{tramer2016stealing}, where the authors showed that it is possible to fully reconstruct a linear binary classifier by using enough queries to allow the model parameters to be computed by solving a system of equations. Using randomly generated thief datasets, they also proposed approximate attacks against decision trees, shallow neural networks, and SVMs.

Other learning attacks test the use of thief datasets that are related to the original domain or not. The CopyCat attack~\cite{correia2018copycat} showed that it is possible to steal a convolutional network that performs image classification using both types of thief datasets. The Knock-off attack~\cite{orekondy2019knockoff} proposed the use of reinforcement learning to select samples from a thief dataset that would make the attack as query efficient as possible. The attack was tested against image classifiers and used both types of thief datasets to construct the thief dataset.

A model extraction attack shares many similarities with active learning, where there is an oracle function that provides labels for each sample when queried. Labeling is usually a costly function and may or may not involve a human. The active learner queries the oracle to get labels for the samples, which are then used to train a model for a given task. Defining a strategy to select the best samples that make the learning as efficient as possible is part of the active learning algorithm. 

Similarly, in model extraction attacks, the target model plays the part of the oracle, and the attacker aims to learn a good approximation of the oracle function. This connection of model extraction to active learning was explored in~\cite{chandrasekaran2020exploring}, where the authors proposed using query synthesis techniques to extract different types of ML models such as decision trees, random forests, SVMs, and linear models. The use of synthetic samples had been previously proposed in~\cite{papernot2017practical}, where a small initial seed of real images was used to create new images using Jacobian-based dataset augmentation, and these synthetic data were used to query the target model.  

The ActiveThief attack was proposed in~\cite{pal2020activethief}, where the authors tested different active learning sampling strategies to create query-efficient attacks against image and text classifiers. They also proposed to generate adversarial samples for the data in the thief dataset and then use the samples with the smallest distance to their respective adversarial samples as a sampling strategy. The use of adversarial attacks was also proposed in "CloudLeak"~\cite{yu2020cloudleak} with the difference that the synthetic adversarial samples were used to train the surrogate model. The attack targets image-based classifiers that are deployed in the cloud. One of the problems in using synthetic in the malware domain is that binary files are much harder to construct than images.

The use of semi-supervised learning in model extraction attacks was proposed in~\cite{jagielski2020high}, which is a promising avenue for learning-based attacks. The authors proposed a model extraction on 2-layered neural networks with ReLU activations in the same paper. Full extraction attacks on neural networks have also been proposed either by the use of side-channels~\cite{batina2019csi} or by performing cryptanalysis attacks~\cite{carlini2020cryptanalytic}. While these attacks can recover the actual neural network weights, they are limited to specific models and architectures.

Defenses against model extraction attacks include modification of the output vectors of the target model, restricting model output information, and watermarking. Changing the output confidence vectors of specific inputs aims to make model stealing harder, either by adding noise or perturbations to the output~\cite{tramer2016stealing,lee2019defending} or by restricting the amount of information by providing complex labels instead of a confidence vector~\cite{jagielski2020high}. However, output perturbations come with a trade-off between model utility and defending the target model. Watermarking defenses are designed to make the model overfit specific outlier input-output pairs of data~\cite{adi2018turning,darvish2019deepsigns,nagai2018digital,zhang2018protecting}. The defender then performs a verification process where they show that the stolen model has learned to respond to these outliers. However, this defense assumes that the attacker will make queries containing data samples close to the outliers. More recent watermarking defenses overcome this issue by proposing ways to generate watermarked data that push the model to learn representations that are "entangled" with those of the original training data~\cite{jia2020entangled}. Another proposal is to use an additional surrogate model that uses the original training data and perturbed outputs of the original model for its training~\cite{chakraborty2022dynamarks}. The surrogate is used to respond to user and attacker queries. During verification, they compare the stolen model outputs with that of the target and the surrogate model and decide. The watermarking methods assume that the target model owner has query access to the stolen model and that they can perform a verification process. This assumption is not valid against attackers without the intention to publish their models. In addition, some watermarking attacks are based on manipulating the input data and inserting a watermark. While this is possible in image data, it is not a straightforward process in binary file modifications. 

Generating adversarial samples for the evasion of malware classifiers is a less researched topic than adversarial attacks in more traditional AI domains such as computer vision~\cite{demetrio2020adversarial}. Some of the works focus on the generation of adversarial Windows malware in a black-box setting where the adversary does not know the model parameters~\cite{anderson2018learning, castro2019aimed, ceschin2019shallow, demetrio2021functionality, fleshman2018static, song2021mabmalware}. 

Two black-box attack papers propose to use reinforcement learning to choose modification actions that evade the classifiers~\cite{anderson2018learning, song2021mabmalware}. Two other articles suggest choosing binary modifications using genetic programming~\cite{castro2019aimed, demetrio2021functionality}. In addition to these black-box attacks, Ceschin et al.~\cite{ceschin2019shallow} showed that implementing a dropper is sufficient to bypass malware classifiers and AVs, and Fleshman et al.~\cite{fleshman2018static} proposed a framework that performs random modifications to malware to evaluate the robustness of classifiers and AVs.

Other work that does not require white-box access to a machine learning model either assumes access to the confidence output of the classifier (soft-label)~\cite{labacacastro2021universal} or proposes the creation of surrogates~\cite{rosenberg2018generic} or uses independent models~\cite{huang2019malware, rosenberg2020generating}. These surrogate models would allow them to perform white-box attacks as a next step. Huang et al.~\cite{huang2019malware} trained a model using a different dataset than the one used for training the target model. Then they used the first model to generate adversarial malware. The adversarial malware were generated in the feature space (API calls) and not by modifying the binary files. The test set was gathered independently from the training dataset, but the details are sparse. 
 
Rosenberg et al.~\cite{rosenberg2018generic} also worked with API calls from Windows binaries. They created a surrogate model using a Jacobian-based augmented~\cite{papernot2017practical} dataset, which was then used for white box adversarial sample generation. The Jacobian dataset augmentation required an initial set of real data taken from the test set. Finally, Rosenberg et al.~\cite{rosenberg2020generating} trained additional models using subsets of the Ember dataset~\cite{anderson2018ember} in different variations in the number of features and the overlap between the training sets. Using explainability algorithms, they identified the most helpful features for generating malware that can evade the target model. It has to be noted that only~\cite{rosenberg2018generic} created surrogates that retrieve labels from the target model, i.e., used a model stealing attack. The other two works based their attacks on transferability between models trained on similar datasets.

Both~\cite{rosenberg2020generating} and \cite{rosenberg2018generic} modify binaries in an end-to-end fashion, i.e., aiming at generating functional binaries, which is also the case with the fully black-box attacks. A subset of the papers tests the validity of their generated binaries~\cite{song2021mabmalware, castro2019aimed} or checks the malware against a real antivirus or VirusTotal~\cite{anderson2018learning, ceschin2019shallow, demetrio2021functionality, fleshman2018static, song2021mabmalware}.

Finally, the focus on black-box attacks is motivated by the fact that antivirus products are complex black boxes that are given an input (malware). The output is the result of the detection process. However, it is possible to discover information about their internals using reverse engineering techniques~\cite{mougey2021blackhat, koret2015antivirus}. Reverse engineering can be time-consuming, and it is aimed at each product. The authors of AVLeak~\cite{blackthorne2016avleak} proposed using a fingerprinting method to detect AV emulators. In~\cite{quarta2018toward}, the authors proposed a methodology that aims to discover whether an antivirus implements emulation techniques, whether they perform unpacking, and whether they rely on heuristics for its detections. Sometimes, the AVs themselves provide too much information via their logging mechanisms that allow attackers to bypass them~\cite{ashkenazy2019bsides}.
\section{Threat Model}
\label{sec:threat-model}

\begin{figure}[!t]
\centering
 \includegraphics[width=0.60\textwidth]{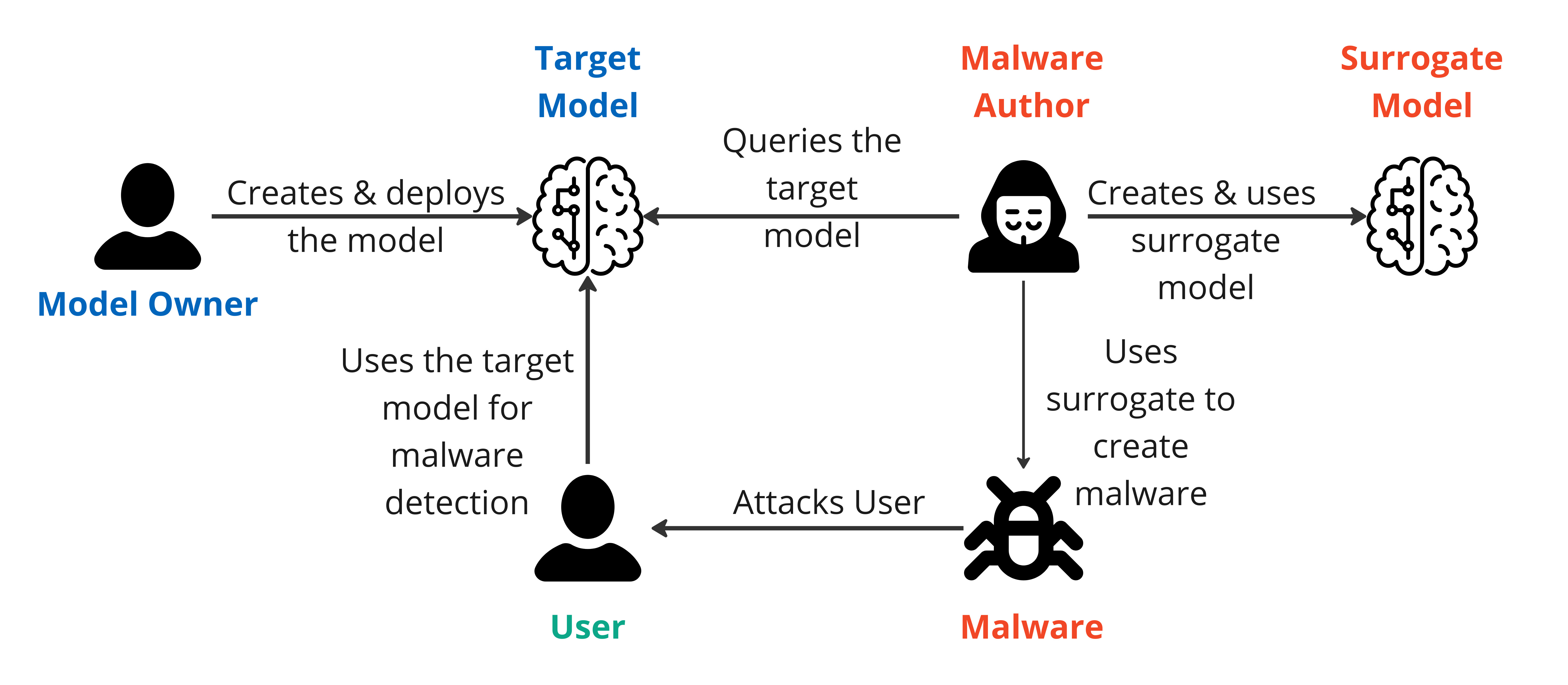}
\caption{Threat model}
\label{fig:threat_model}
\end{figure}
The idea behind model extraction attacks is to generate a surrogate model that behaves similarly to the model under attack. In the malware detection and classification domain, we can see two main adversaries: first, the adversaries interested in evading the model with adversarial malware, and second, those interested in stealing the model to resell it in the market. The first type of attacker is interested in fidelity attacks, while the second type cares more about task accuracy attacks. This paper focuses on the first attacker. The attack needs to be relatively fast to survive updates on the part of the AVs and should be feasible with relatively small datasets. 

Figure~\ref{fig:threat_model} is a graphical depiction of our threat model. The model owner creates and deploys the target model the end user uses for malware detection. The malware author is the adversary who wants to steal the target model to create a surrogate model. In order to achieve that, they query the target model and use the query results to train their surrogate. Then the surrogate creates malware that bypasses the user's detection attempts.

Regarding the adversary's capabilities, we assume they can query the models or the AVs during the inference phase, i.e., after the model owner has trained and deployed the target model. The attacker has only black-box access to both models and AVs, meaning that they can scan a binary file and get back a hard label (benign / malware). We consider having access to a soft label (a probability or confidence vector) a gray-box attack. While several model extraction papers in the literature assume a gray-box attacker, we argue that this is not a realistic scenario for malware detection. To our knowledge AV products do not provide confidence outputs for their detections and creating a surrogate model is a way to obtain those confidence vectors.

The attacker also has limited knowledge of which feature space the target model may be using. For the stand-alone machine learning models we evaluated, the features are known (Ember v2 features), but they are not known in the case of AVs. Similarly, the architecture and model types of the classifiers are known in the case of stand-alone models. However, there is almost no information on what AV companies may be using.

Regarding the knowledge about the datasets used for training the target models, the attacker knows the training sets used for the stand-alone Ember and Sorel20m models; but they do not know the training data of the AVs. 

\section{Methodology}
\label{sec:methodology}

\begin{figure}[!t]
\centering
\includegraphics[width=0.60\textwidth]{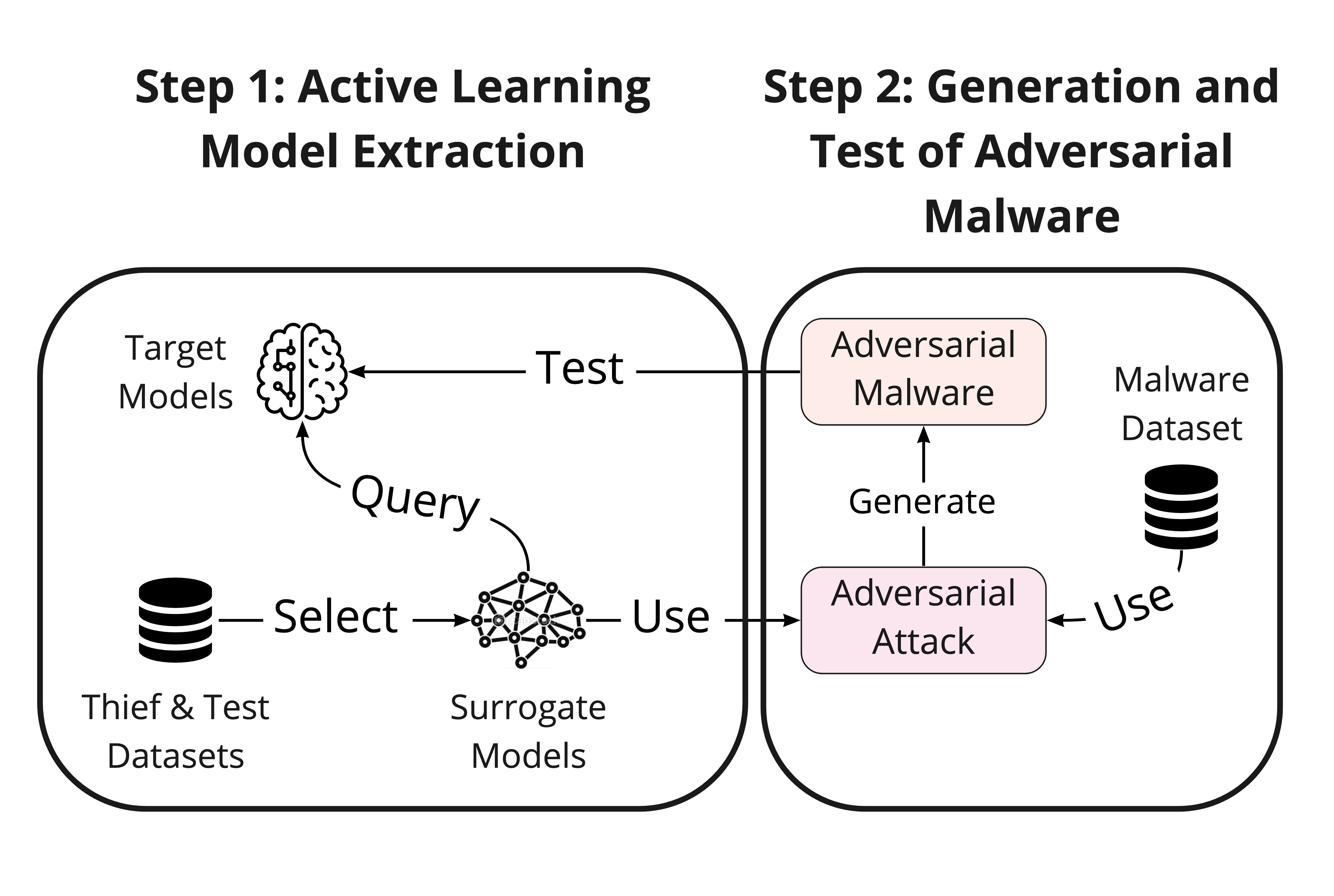}
\caption{High level methodology of our work: The first step creates surrogate models that behave similarly to the target models. The second stage uses surrogates to create adversarial binary malware that evade target models.}
\label{fig:methodology}
\end{figure}

A high level diagram of the attack methodology is shown in Figure~\ref{fig:methodology}. Playing the part of an adversary, the first part of the methodology creates surrogate models that \textit{extract}, or \textit{steal}, the functionality of the target models. This part of the attack is explained in Subsection~\ref{sec:model-extraction-attacks-methodology}. The second part of this paper uses those surrogate models to create adversarial malware binaries that test their evasion ability against target models, including real antivirus systems (both offline and online). This second part of the methodology is explained in Subsection~\ref{sec:adversarial-malware-generation-methodology}.

\subsection{Model Extraction Attack Methodology}
\label{sec:model-extraction-attacks-methodology}
We extract the functionality of target models by creating surrogate models that closely resemble the performance of the target models. The goal of a surrogate model is not to have good detection performance but to have a performance close to the target model, using the \textit{agreement} metric to evaluate the attack. 

\subsubsection{Active Learning Model Extraction}
\label{sub:AL-model-extract}
\begin{figure}[!t]
\centering
 \includegraphics[width=0.60\textwidth]{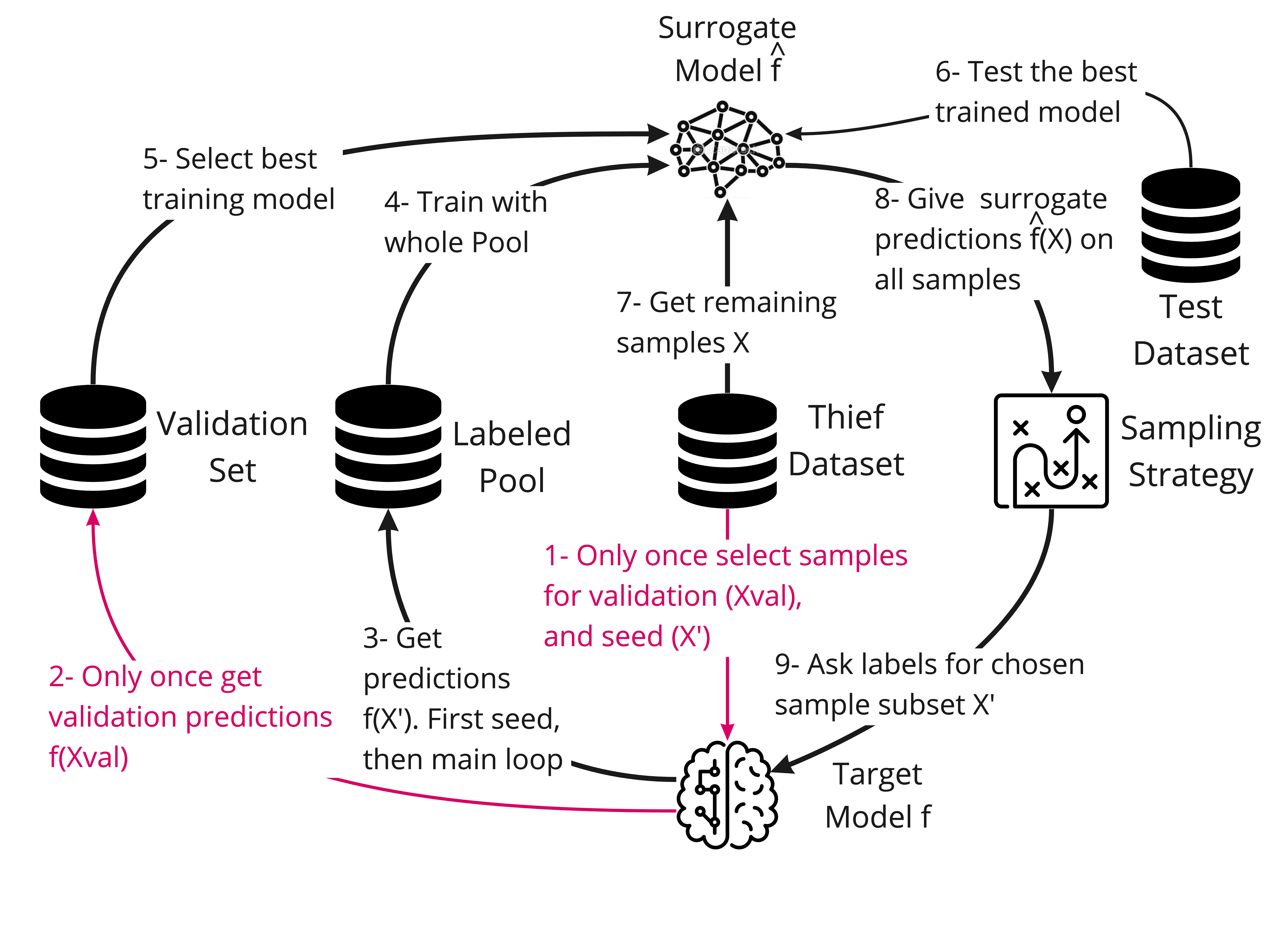}
\caption{Active Learning model extraction algorithm to create surrogate models $\hat{f}$. Querying target models $f$ with a subset selection of the thief dataset. All labeled samples are added to the Labeled pool that is used to train the surrogate. The main loop consists of steps 3,4,5,6,7,8,9; steps 1 and 2 (in red) are performed once.}
\label{fig:active_learning_schema}
\end{figure}

Our surrogate models are trained using an Active Learning approach~\cite{activelearning}, similar to the ActiveThief framework~\cite{pal2020activethief}. We use active learning because it is suitable for fidelity extraction attacks, our primary goal. In addition, these attacks can work with label-only outputs, and they make no assumptions about the type of target or surrogate models. The Active Learning approach schema is shown in Figure~\ref{fig:active_learning_schema}. It uses two main datasets, the \textit{thief dataset} that is used to query the target model and the \textit{test dataset} that is used for attack evaluation. 

Figure~\ref{fig:active_learning_schema} shows that the first step of the attack is to select validation randomly and seed samples from the thief dataset (step 1), which is done only at the beginning of the process. The validation samples are used to query the target model and retrieve their labels. The data samples and their respective target labels are stored in the validation set (step 2). Similarly, the seed data labels are also retrieved from the target model and are added to the Labeled Pool of samples (step 3). In step 4, the complete Labeled Pool is used to train the surrogate model, and during training, the validation set is used in each round to select the best-trained model (step 5). In step 6, the trained surrogate model is tested using the test dataset to obtain metrics for the training round. The remaining samples of the thief dataset are used to retrieve a confidence value from the surrogate model (step 7). The sampling strategy uses the confidence values to choose a subset of the thief dataset (step 8) to send to the target model for labeling (step 9). The target model labels the data that are then added to the Labeled Pool (step 3). The process is repeated for several rounds. 

An essential part of the methodology is the \textit{query budget}, which limits the number of queries allowed on the target model, given the cost of the queries and potential security limitations. We use and separate 20\% of the query budget for the validation samples and 10\% for the seed samples. The rest of the budget is split equally for all the rounds for the query samples.

\subsubsection{Sampling strategies}

Figure~\ref{fig:active_learning_schema} shows that a sampling strategy is used to select samples after the surrogate model generates the predictions for the whole thief dataset. In each round, the chosen strategy is run to select a subset of samples from the predicted thief dataset.

The sampling strategy aims to find those samples that would provide the most value for the next training iteration when asking the target model to label them. These samples are expected to better map the decision boundary for the surrogate model. Good sampling strategies may improve the surrogate model's performance and decrease the number of queries used from the query budget. The sampling strategies used and compared were:

\begin{enumerate}
    \item \textbf{Random sampling} selects $n$ samples randomly, following a uniform distribution. 
    \item \textbf{Entropy sampling} selects the top $n$ samples with the highest Shannon entropy calculated over the predictions.
    \item \textbf{Entropy$+$k-medoids} selects a subset of 10,000 samples from the thief dataset based on entropy (as previously explained) and then splits the subset into $k$ clusters using k-medoids. The $k$ centers are then selected as query samples. The motivation was that the entropy sampling provides data samples close to the decision boundary but not necessarily diversified. This strategy is a contribution of our paper and is inspired by the "entropy+k-center" algorithm~\cite{pal2020activethief}. A major difference is that they perform the k-center algorithm over the prediction vectors while we use k-medoids over the data points.
    \item \textbf{MC dropout+Entropy} is a combination of Monte Carlo dropout~\cite{gal2016dropout, beluch2018power} and entropy sampling. First, we use MC dropout with 20 neural network forward passes to retrieve prediction vectors for the thief dataset. Then we perform entropy sampling on the vector averages. MC dropout has not been used before for model stealing attacks, and it is tested for the first time in this paper. 
\end{enumerate}

\subsection{Adversarial Malware Generation Methodology}
\label{sec:adversarial-malware-generation-methodology}
The second part of our research evaluates the possibility that an attacker can use surrogates to perform a second-stage attack. In this case, the attacker can use the surrogates to create adversarial malware with the expectation that if they evade the surrogates, they will be able to evade being detected by the targets. For this purpose, we use two different attacks, MAB~\cite{song2021mabmalware} and GAMMA~\cite{demetrio2020adversarial}.

The MAB attack is based on reinforcement learning. It works in two stages: first, it creates an adversarial binary, and second, it minimizes the modification of the binary. In the first stage, it selects a series of actions that modify the malware binary to evade a given target. These actions are selected from a set of actions such as appending benign sections or content to a binary, removing certificates, removing debug information, etc. In the second stage, it iteratively removes the actions previously added to find the \textit{minimal} binary that is still evasive. All the actions applied to the binary retain its functionality.   

The GAMMA section injection attack was implemented in the \textit{secml\_malware} library~\cite{demetrio2021secmlmalware}. The GAMMA attack adds benign sections into malicious PE files without compromising its functionality, under the assumption that these benign sections may help evade detection. The injected sections modify several features that may be relevant to malware classification, such as the number of sections, byte histograms, and strings. However, some features are not affected, such as certificate or debug data information. GAMMA searches for the optimal benign sections whose content and location minimize the confidence of the target model. This search is done using genetic algorithms.

\subsection{Performance Metrics}
\label{sec:performance_metrics}

The most important metric to measure the quality of a surrogate model, in the fidelity type of attacks, is the agreement between the surrogate and the target model, that measures how many similar predictions the two models have on the test set $\mathcal{D}_{test}$. 
\begin{equation*}
    Agreement(f, \hat{f}) = \frac{1}{|X_{test}|} \sum_{x \in X_{test}} \mathbbm{1}(f(x) = \hat{f}(x))
\end{equation*}
where $f(x)$ and $\hat{f}(x)$ are the prediction labels on the sample $x$. Accuracy as a metric is more suitable for task accuracy extraction attacks, however we measure it as well:
\begin{equation*}
    Accuracy(\hat{f}) = \frac{1}{|\mathcal{D}_{test}|} \sum_{x,y \in \mathcal{D}_{test}} \mathbbm{1}(\hat{f}(x) = y)
\end{equation*}

Both agreement and accuracy are measured at fixed FPR levels, usually 0.01 or lower, depending on the target. For the adversarial malware generation experiments, we used the TPR or detection rate as a metric, which measures the number of malicious binaries detected as malicious by the classifier or AV.

\section{Combining Transfer and Active Learning}
\label{sec:combining-tl-al}
Active learning model extraction attacks (Subsection~\ref{sub:AL-model-extract}) aim to steal a target model using as few queries to the target as possible. However, for some types of surrogate models, such as neural networks, using these few queries to train them may not be optimal. If an attacker has access to additional data with ground-truth labels, they may use them to stabilize the training process and improve the surrogate model. This additionally labeled dataset, therefore, would not be used to query the target. The additional dataset must not interfere with learning the target's decision boundary. Therefore a sample weighting process is necessary. 

More formally, transfer learning (TL) or domain adaptation is the process where knowledge from a source domain $\mathcal{S}$ where we usually have plentiful labeled instances is transferred to a target domain $\mathcal{T}$ where labeled data are usually scarce. Data-based approaches view TL as knowledge transfer via data transformation and adjustment~\cite{zhuang2021survey}. One of the data-based approaches is instance-based TL, where the source domain data samples supplement the training data of the target domain using some form of weighting in the loss function. A general formulation for this approach is the following:

$$ \mathcal{L}(X, y; \theta) = \alpha \mathcal{L}_{T}(X, y; \theta) + (1-\alpha) \mathcal{L}_{S}(X, y; \theta)$$ where $\alpha \in [0, 1]$ and $\mathcal{L}_{T}$ and $\mathcal{L}_{S}$ are the losses over the source and target domain data instances, respectively~\cite{blitzer_learning_2007}.

In the above formulation, an active learner can play the part of the target domain, where an oracle provides the labels for the target data samples. This makes active learning more efficient since it can take advantage of more labeled instances from the source domain and thus minimize the queries of the target domain. In the literature, different proposals aim to select the most suitable data from the source domain. However, in our case, we use all available data.

To adapt the above to the model stealing methodology from Section~\ref{sec:model-extraction-attacks-methodology}, all is required is to add another labeled source of data $\mathcal{D_{S}}$, to the labeled pool of data in Figure~\ref{fig:active_learning_schema}. The active learning sampling strategies work exactly the same as before, using the thief dataset $\mathcal{D}_{thief}$. During training, we perform sample weighting using the loss function below:

$$ \mathcal{L}(X, y; \theta) = \frac{1}{s+t} (\sum_{i=1}^{s}w_{i}\mathcal{L}(x_{i}^{(S)}, y_{i}^{(S)}; \theta) + \sum_{i=s+1}^{s+t}w_{i}\mathcal{L}(x_{i}^{(T)}, \hat{y}_{i}^{(T)}; \theta))$$

where, $s$ is the number of source data points, $t$ is the number of labeled thief data points using the AL queries, and $w_i$ is calculated as follows:

$$ w_i =
  \begin{cases}
     \frac{(1-\alpha)(s+t)}{s}      & \quad \text{if } i \leq s \quad (x_i \in \mathcal{D}_S)\\
    \frac{\alpha(s+t)}{t}    & \quad \text{if } s < i \leq s+t  \quad (x_i \in \mathcal{D}_{thief})
  \end{cases}$$

The $w_i$ is formulated in a way that it does not only contain the $\alpha$ but, in addition, adjusts the weights based on the number of training data points in the $\mathcal{D}_{S}$ and $\mathcal{D}_{thief}$ respectively. The values of $\alpha$ determine the trade-off between the source and target domain data. When $\alpha=1$, we only use the target data ($\mathcal{D}_{thief}$), and this is essentially the same as performing the original active learning model extraction attack. When $\alpha=0$, we only use the source data, similar to using an independent model as a surrogate. According to~\cite{ben-david_theory_2010}, the optimal $\alpha$ provides a measure of the similarity of the source and target domains. If the optimal $\alpha$ is close to zero, the two domains are so similar that it is unnecessary to use many of the target data points. Similarly, if the optimal value of $\alpha$ is high, the two domains are quite dissimilar, and therefore using $\mathcal{D}_{S}$ might not be that beneficial. In all our experiments, we set $\alpha=0.8$. The value was determined empirically by testing different values of $\alpha$ on different targets.

The combined attack described in this section assumes that the attacker has access to another labeled dataset ($\mathcal{D}_{S}$), however, this dataset is not used for making queries to the target. Therefore, this allows the attacker to use publicly available data such as Ember~\cite{anderson2018ember} and Sorel20m~\cite{harang2020sorel20m}, usually coming as pre-extracted features and do not contain the original binary files. Such datasets cannot be used to query AV targets requiring actual files for scanning. However, our combined transfer and active learning attack allows the attacker to combine smaller datasets of actual files with larger feature-only datasets.

\section{Surrogate Models}
\label{sec:surrogate_models}

\subsection{Baseline Surrogate Architectures}
\label{sec:baselines}

The two \textit{baseline} models that were used in our experiments are: (i) a LightGBM (LGB) similar to the Ember2018 and Sorel-LGB targets, and (ii) a fully connected neural network (FFNN) that was inspired by the architecture used in~\cite{aloha2019, harang2020sorel20m}.

The LightGBM surrogate uses similar settings as the original Ember2018 model with the most important parameters being the number of leaves ($2048$) and max depth set to $15$. 

The two \textit{baseline} models that were used in our experiments are: (i) a LightGBM (LGB) similar to the Ember2018 and Sorel-LGB targets, and (ii) a fully connected neural network (FFNN) that was inspired by the architecture used in~\cite{aloha2019, harang2020sorel20m}.

The LightGBM surrogate uses similar settings as the original Ember2018 model, with the most important parameters being the number of leaves ($2048$) and max depth set to $15$. 

The FFNN architecture consists of four dense layers with decreasing number of hidden units and a sigmoid activation at the final layer. Each dense layer uses Exponential Linear Unit (ELU) activations~\cite{clevert2015fast} and is followed by a normalization layer~\cite{ba2016layer} and dropout set to $0.3$. The input layer has a size of 2381, corresponding to the number of Ember features used during feature extraction.

In order to derive the final architecture of the baseline FFNN model, we used the Ember training data (Section~\ref{sec:datasets}) to decide on the number of fully connected layers, the size of the layers, the activation functions, the normalization layers, and the dropout value. We used grid search with the following parameters:
\begin{itemize}
    \item Number of fully connected layers: 3 to 6 
    \item Size of the intermediate layers: 1024, 512 and 256
    \item Activation functions: RELU, ELU
    \item Dropout: 0.2, 0.3, 0.4
    \item Normalization: batch, layer normalization, or no normalization layers
\end{itemize}

The architecture was chosen based on the performance of a validation subset (20 \% of the training data) on multiple seeds. The model was tested on both the Ember and the Sorel test sets to verify that the architecture performs well enough on both datasets in low FPR conditions.  

The FFNN requires an extra data pre-processing step, and we used the robust scaler from the scikit-learn library~\cite{scikit-learn} for that purpose. During all experiments, the scaler fit the training set, and the scaling transformation was applied to the test set. Both neural network models used binary cross-entropy as the loss function and Adam~\cite{kingma2015adam} as the optimizer.

\subsection{A New Surrogate Architecture}
\label{sec:new_architecture}
The malware classification problem requires low FPR, so the surrogate models must alsowork better under low FPR conditions. Initial experimentation with the baseline FFNN model showed that the FFNN surrogate was always a worse surrogate than the LGB baseline, especially at low FPR settings. Inspired by the ResNet architecture~\cite{he2016deep} that is designed to stabilize the network optimization, we propose a new deep neural network architecture that uses the additional knowledge of true labels ($y_{true}$) that are readily available to an attacker. True labels are the ground truth actual labels of malware samples. Having these labels is a realistic assumption since an attacker will use all the information at their disposal in the training dataset to improve the surrogate model. Using the true labels as an additional input, we created a model that essentially learns to predict the \textit{target's label} $y_{target}$ given as input both the data $X$ and the true labels $y_{true}$:

$$p(\hat{y}_{target} | X, y_{true}; \theta)$$ where $\theta$ are the model parameters. 

\begin{figure}[!t]
\centering
\includegraphics[width=0.5\textwidth]{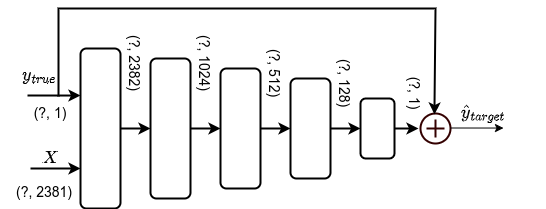}
\caption{Our proposed surrogate dualFFNN architecture using true labels and a skip connection.}
\label{fig:new_architecture}
\end{figure}

Figure~\ref{fig:new_architecture} shows the proposed neural network architecture, called \textbf{dualFFNN}. The main body is the same as the FFNN baseline architecture~\ref{sec:baselines}. The true labels $y_{true}$ are concatenated with the input data $X$ in the input layer that has a size of 2382. They also form a skip connection with the final layer of the model. The skip connection allows the model to learn the difference between the true and target labels and provides stability during training.

The dualFFNN architecture may have some limitations as a surrogate if the target model labels are always correct. However, this is not a case that we encountered when dealing with a large number of targets. In addition, the dualFFNN cannot be used as a malware detector because it requires the true label of an input sample. However, the purpose of this study is not to create surrogates that detect malware but surrogates that emulate the decision boundary of a given target. Finally, even though this architecture is designed for black-box attacks, one of the ways the dualFFNN surrogate could be used in a white-box attack is to treat it as an intermediate labeling function. The dualFFNN model can learn to label another much larger dataset like the target with only a few thousand queries to the actual target. This much larger dataset could subsequently be used to train another model that can be used in white-box attacks.

\subsection{Model Trained Using Transfer and Active Learning}
The second surrogate model used in our experiments is a fully connected neural network (FFNN) with the same main body as the baseline FFNN (Section~\ref{sec:baselines}) trained using the combined transfer and active learning approach introduced in Section~\ref{sec:combining-tl-al}. This model does not use the skip connection of the dualFFNN. However, it requires a labeled source dataset. In the remaining sections, we refer to this model as \textbf{FFNN-TL}.

\section{Models Under Attack}
\label{sec:models-under-attack}
All model extraction attacks are performed on three stand-alone classifiers and four commercial AVs. The stand-alone classifiers are:
\begin{itemize}
    \item \textbf{Ember2018} is a Gradient Boosting Tree model based on LightGBM~\cite{ke2017lightgbm} that is part of the Ember 2018 dataset.
    \item \textbf{Sorel-FFNN} is one of the two models distributed as part of the Sorel20m~\cite{harang2020sorel20m} dataset. It is a fully connected neural network trained using the dataset mentioned above. 
    \item \textbf{Sorel-LGB} is a LightGBM model that is also distributed as part of the Sorel20m dataset~\cite{harang2020sorel20m}.
\end{itemize}

The four \textbf{AVs} were installed in Windows 10 virtual machines (VMs). They were among the top-scoring AVs in "The best Windows antivirus software for home users"~\footnote{https://www.av-test.org/en/antivirus/home-windows} and were chosen because they publicly disclosed that they use machine learning methods.

All three stand-alone ML models use the Ember v2 feature set, which contains information related to static features. While the ML models provide confidence vectors, we only use the classification labels ($0$ for benign / $1$ for malware), calculated using a threshold of $0.8336$ for Ember and $0.5$ for the Sorel20m models (See Subsections~\ref{sec:stealingember} and \ref{sec:stealingsorel} for more details on the selection of the threshold values). On the other hand, the AVs are complete black boxes, only concluding if a file is malicious or not. Each AV is asked to scan the binary file without executing the file. Some AVs provide more configuration options than others, but we mostly used their default settings.

\section{Datasets}
\label{sec:datasets}

In this work we used three main datasets: the Ember 2018~\cite{anderson2018ember}, the Sorel20m~\cite{harang2020sorel20m}, and an "internal" dataset of binary files that was required for the attacks against the Avs. Ember and Sorel20m were used in the attacks against the standalone models.

Ember is a dataset that consists of extracted features from Windows Portable Executable (PE) files~\cite{anderson2018ember}. Ember 2018 is the second version of the dataset (from now on, Ember dataset), which contains data extracted from binaries that were first seen during 2018, split into training and test sets. The training set consists of 300,000 clean samples, 300,000 malicious samples, and 200,000 "unlabeled" samples. The test set contains 100,000 clean samples and 100,000 malicious samples. Each sample has 2,381 static features related to byte and entropy histograms, PE header information, strings, imports, data directories, etc. The unlabeled samples were truly unlabeled in the first version of the dataset and they were not used to train or validate the Ember model. However, the second version of the dataset included an additional "avclass" label with a generic AV class for each malicious sample, even the unlabeled ones. So in the unlabeled dataset, if a sample has an "avclass," we take it as "malicious." We consider it "benign" if it does not have an "avclass" label. In this work, we used the unlabeled part of the dataset as the \textit{thief dataset} for the model extraction attacks of the Ember target model and the test dataset for testing the surrogate model performance.

The Sorel20m dataset~\cite{harang2020sorel20m} was released in 2020 and contained pre-extracted ember features for 20 million benign and malicious binaries. It also contains all malware binary files in a deactivated form. However, no benign binary files were included, making it hard to use the dataset when attacking AVs. A subset of the validation set was randomly sampled (200,000 samples) and was used as a thief dataset for Sorel20m targets. A subset of the test set (200,000 samples) was randomly sampled and used as the test set against the Sorel20m targets.

Ember 2018 provides only the extracted features of the PE files and not the binaries themselves. Sorel20m contains modified malware binaries but not benign files. When stealing AV functionality, it is required to query the AVs using actual files. Therefore, we used an internal curated dataset of malware files that were first seen mostly between the years 2018-2020. The top 10 families in terms of samples from that dataset were used as malicious samples and the total number of malware files was 140,288.


\begin{table}[!htp]\centering
\caption{Number of samples in the all datasets including train/test splits.}
\label{tab:internal_dataset}
\begin{tabular}{lrrrrr}
\toprule
Dataset & &Thief &Test \\
\midrule
\multirow{2}{*}{Ember} & Malware & 103,577 & 100,000\\
& Benign & 96,433 &  100,000 \\
\midrule
\multirow{2}{*}{Sorel20m} & Malware & 135,131 & 122,771 \\
& Benign & 64,869 & 77,229 \\
\midrule
\multirow{2}{*}{Internal} & Malware & 116,192 & 30,482 \\
&Benign & 24,096 & 12,130 \\
\bottomrule
\end{tabular}
\end{table}

We obtained the benign binaries of the internal dataset by taking all the PE files after a clean installation of virtual machines running Windows 10 (64-bit) and Windows 7 (32-bit) and after installing well-known verified software. The criteria to choose the software were: a) it must come from well-known, reliable sources, b) it should be used in various tasks such as office tasks, software development, entertainment, security analysis, etc. After the installation, we retrieved all .exe and .dll files, calculated their SHA256 hashes, and extracted their Ember v2 features. This process gave us approximately 20,000 unique benign binaries. In addition, we used the Virus Total platform and retrieved .exe and .dll binaries with the tag "trusted" that were first seen before April 2021. The query used for this purpose was \textit{"tag:trusted and (tag:pedll or tag:peexe) and fs:2020-04-01T00:00:00- and not tag:assembly"}. The use of the tag "trusted" was suggested by VirusTotal engineers. The total number of clean samples was 36,226.

The internal dataset was split into thief and test sets based on the timestamp of 'first seen' retrieved from Virus Total for the clean files. The cut-off date was 31.08.2019, and all binaries with earlier timestamps were placed in the thief dataset, while the rest were part of the test set. The exact number of benign and malware samples can be seen in Table~\ref{tab:internal_dataset}.

\section{Stealing Stand-Alone ML Models}
\label{sec:stealing_stand_alone_ml_models}
The creation of surrogates is sometimes proposed in the adversarial malware attacks literature~\cite{rosenberg2018generic, huang2019malware, rosenberg2020generating}. However, these attacks do not focus on the surrogate creation strategies and do not perform any comparison between different techniques. The first goal of our experiments is to test the agreement of surrogates with their respective target models at low FPR levels using various active learning strategies. The second goal of our experiments is to answer questions about attack transferability and whether or not the type of target and surrogate models is an essential factor in attack efficacy. Finally, an important variable of model extraction attacks is the thief and test datasets used for creating the surrogates and their potential overlap. We try to measure the performance impact of using datasets with different degrees of overlap and coming from potentially different data distributions.

\subsection{Experimental Setup}
\label{sec:experimental-setup}
We tested the four surrogate models described in Section~\ref{sec:surrogate_models} using the following sampling strategies: \textit{random sampling}, \textit{entropy sampling}, and \textit{entropy+k-medoids}. The \textit{MC-dropout+entropy} strategy was tested only for the neural networks. 

The total query budget was set to a maximum of 25,000 queries which is approximately 4\% of the size of the Ember. This was deemed enough for the model stealing attacks against the Ember model since multiple tests with randomly sampled subsets of the Ember thief dataset, showed that the models' performance stabilizes before reaching the 25,00 queries. The same number was also used for the Sorel20m model attacks in order to be able to perform some comparisons. The total number of query rounds was set to 10, allowing 1,750 new samples to be used per query round. 

Each surrogate  model was trained from scratch at each training round using all data in the labeled pool. The FFNN, dualFFNN, and FFNN-TL models were trained for up to $100$ epochs. The LGB model settings were $500$ boosting rounds. The validation set was used for training model selection: the early stopping rounds for the LGB were set to 60, while the \textit{patience} parameter for the neural networks was set to 30. All neural networks were trained using model checkpoints where the best model, based on validation accuracy, was saved and used for testing. The number of training epochs for the neural networks were selected using multiple runs of model training with random subsets of the thief datasets (25,000 data points) and observing the training and validation losses. The number of boosting rounds for the LGB surrogate were selected in a similar manner.

The use of the different datasets in the attacks described in this section is summarized in Table~\ref{tab:datasets_experiments}.
\begin{table*}
\caption{Use of the different datasets in model extraction experiments.}
\label{tab:datasets_experiments}
\begin{tabular}{l | cccc}
\textbf{Target} & $\mathcal{D}_{S}$ & $\mathcal{D}_{thief}$ & $\mathcal{D}_{test}$ & Surrogate models \\
\toprule
\multirow{2}{*}{ Ember2018 (Sec. \ref{sec:stealingember}) } & - & Ember & Ember (test) & dualFFNN, FFNN, LGB \\
& Sorel20m & Ember & Ember (test) & FFNN-TL \\
\midrule
\multirow{2}{*}{ Sorel-LGB, Sorel-FFNN (Sec. \ref{sec:stealingsorel})} & - & Sorel20m & Sorel20m (test) & dualFFNN, FFNN, LGB \\
& Ember & Sorel20m & Sorel20m (test) & FFNN-TL \\
\midrule
\multirow{2}{*}{ Sorel-LGB, Sorel-FFNN (Sec. \ref{sec:attacking-diff-distribution})} & - & Ember & Ember (test) & dualFFNN, FFNN, LGB \\
& Internal & Ember & Ember (test) & FFNN-TL \\
\midrule
\multirow{2}{*}{ AVs (Sec. \ref{sec:stealing-av-functionality})} & - & Internal & Internal (test) & dualFFNN, FFNN, LGB \\
& Ember & Internal & Internal (test) & FFNN-TL \\
\bottomrule
\end{tabular} 
\end{table*}

\subsection{Stealing Ember2018 Model}
\label{sec:stealingember}
The unlabeled part of the Ember 2018 dataset was used as the thief dataset, and the Ember 2018 test set was used for evaluating the attack performance against the Ember2018 target model. The different metrics were measured at an FPR equal to $0.01$. The threshold of the target model was set to $0.8336$ to achieve this FPR level. Each experiment was run five times to randomize the initial seed and validation data selection.

\begin{figure*}[t]
\centering
    \begin{subfigure}[b]{0.32\textwidth}
     \centering
      \includegraphics[width=\textwidth]{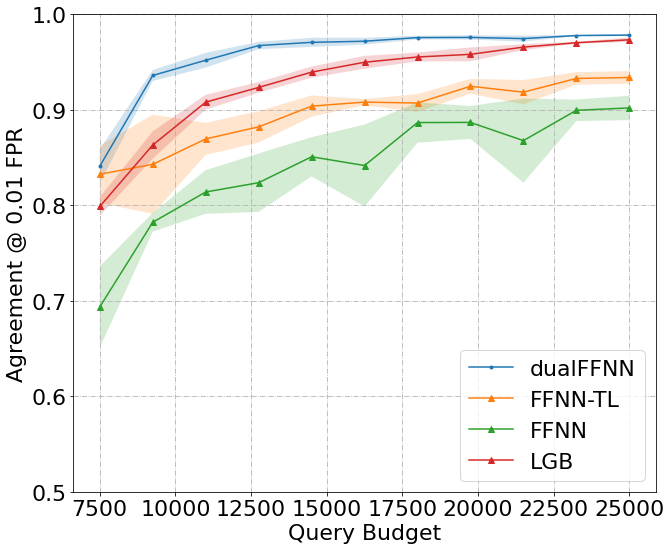}
     \caption{Ember2018.}
     \label{fig:ember_agreement}
\end{subfigure}
\hfil
\begin{subfigure}[b]{0.31\textwidth}
     \centering
      \includegraphics[width=\textwidth]{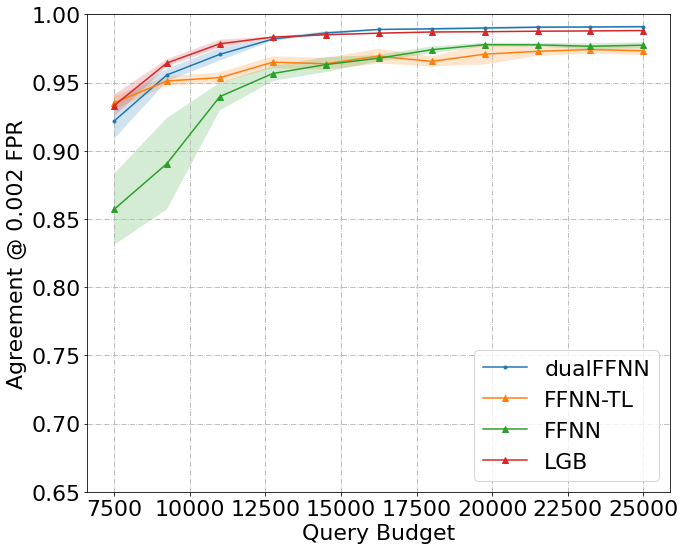}
     \caption{Sorel-LGB.}
     \label{fig:sorel_LGB_agreement}
 \end{subfigure}
 \hfil
 \begin{subfigure}[b]{0.32\textwidth}
     \centering
      \includegraphics[width=\textwidth]{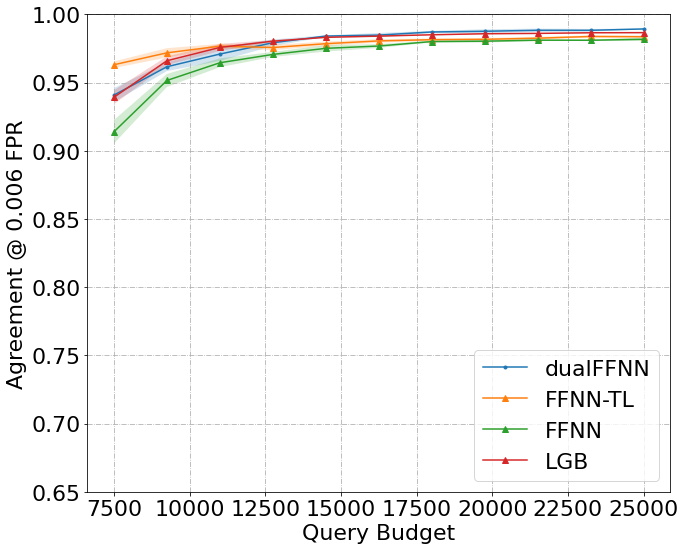}
     \caption{Sorel-FFNN.}
     \label{fig:sorel_FFNN_agreement}
 \end{subfigure}
 \caption{Agreement of selected surrogate models with Ember2018 and Sorel20m target models.}
\label{fig:agreement_ember_sorel}
\end{figure*}

\begin{figure*}[t]
\centering
    \begin{subfigure}[b]{0.32\textwidth}
     \centering
      \includegraphics[width=\textwidth]{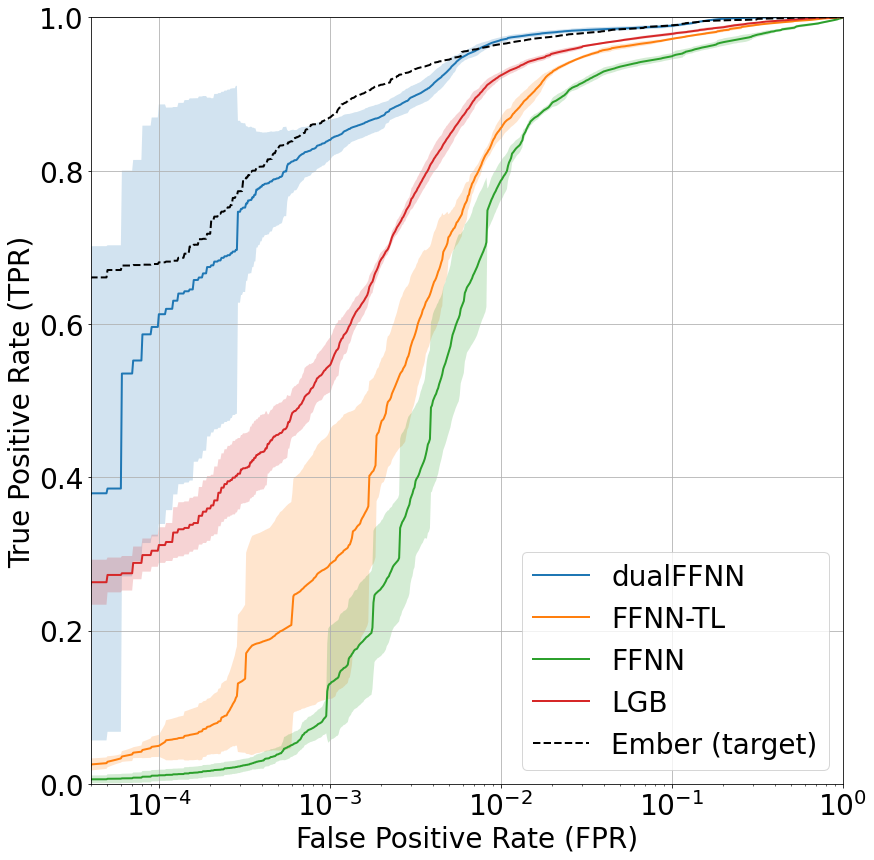}
      \caption{Ember2018.}
     \label{fig:ember_roc}
\end{subfigure}
\hfil
\begin{subfigure}[b]{0.32\textwidth}
     \centering
      \includegraphics[width=\textwidth]{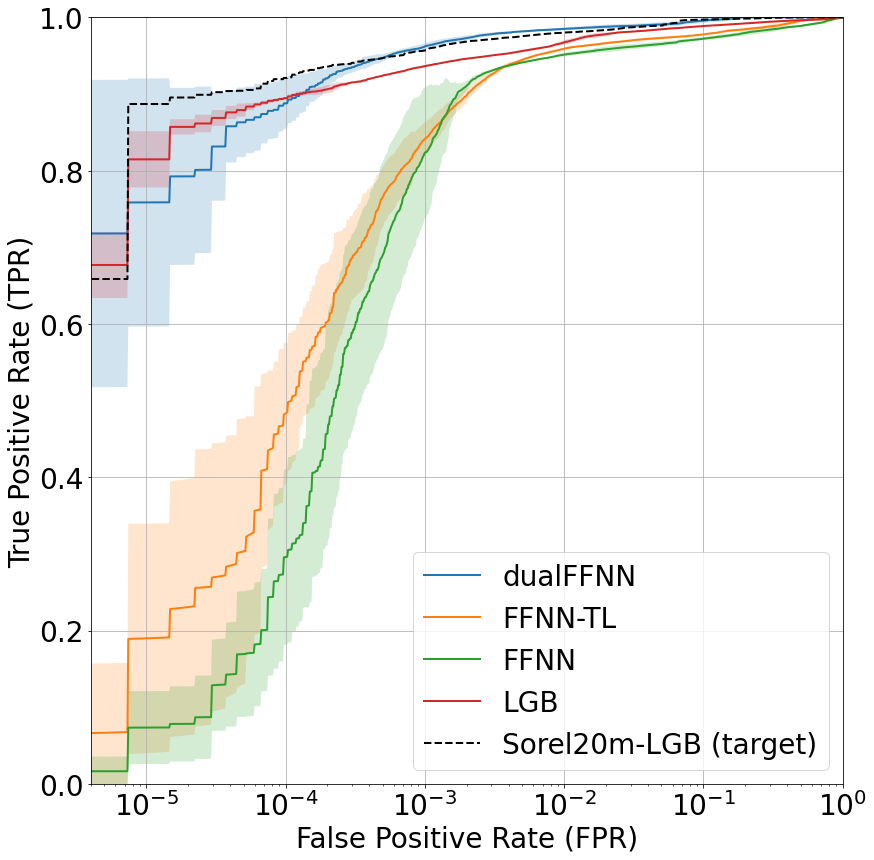}
     \caption{Sorel-LGB.}
     \label{fig:sorel_LGB_roc}
 \end{subfigure}
  \hfil
 \begin{subfigure}[b]{0.32\textwidth}
     \centering
      \includegraphics[width=\textwidth]{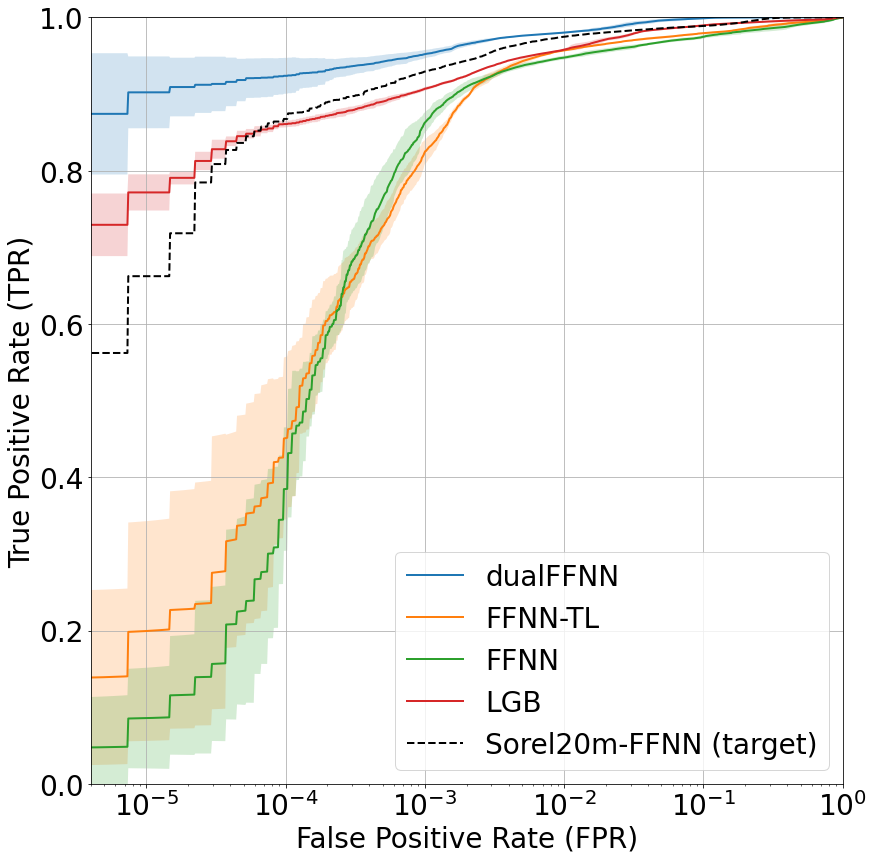}
     \caption{Sorel-FFNN.}
     \label{fig:sorel_FFNN_roc}
 \end{subfigure}
 \caption{ROC curves of selected surrogate models. The dashed line is the ROC curve of the respective target.}
\label{fig:accuracy_ember_sorel}
\end{figure*}

A summary of the results for all strategies is presented in Table~\ref{tab:models_results}. The table shows the agreement and accuracy achieved by each surrogate using 25,000 queries and the threshold required to archive 0.01 FPR. The top line of the table shows the accuracy of the target model, and the lower part shows the results of the surrogates using all the available thief dataset (200,000 data points); therefore no sampling strategy. 

The table shows that, for all surrogates, all non-random sampling strategies outperform the random strategies. Considering only the non-random strategies, the dualFFNN has the highest agreement with the target and has higher accuracy than the target. The LGB surrogates perform closely to the dualFFNN ones, and the LGB, FFNN-TL, and dualFFNN surrogates outperform the FFNN ones.

More importantly, dualFFNN achieved this performance only with a small percentage of the training set (13,000 queries/data points versus 600,000 data points used for training the Ember2018 target), which indicates that choosing the most informative data can be very powerful.

\begin{table*}[!t]
    \caption{Metrics for surrogate models and strategies against the Ember2018, Sorel-LGB and Sorel-FFNN targets using 25K as query budget. The lower part are metrics for the surrogates using the full thief dataset (no sampling) at 0.01 FPR level using 200K query budget.}
    \label{tab:models_results}
    \footnotesize

    \centering
    \begin{adjustbox}{width=\textwidth}
    \begin{tabular}{l c c c | c c c | c c c}

    \toprule
    Target / &  \multicolumn{3}{c|}{\textbf{Ember2018}} & \multicolumn{3}{c|}{\textbf{Sorel-LGB}} & \multicolumn{3}{c}{\textbf{Sorel-FFNN}} \\
    Model \& Strategy & Agreement(\%) & Accuracy(\%) & Threshold & Agreement(\%) & Accuracy(\%) & Threshold & Agreement(\%) & Accuracy(\%) & Threshold  \\
    \midrule
    Target & - & 96.50 & 0.8336 & - & 98.86 & 0.5000 & - & 98.68 & 0.5000 \\
    \midrule
    dualFFNN Random & 95.65 & 95.54 & 0.9835 & 98.14 & 97.92 & 0.9746 & 98.05 & 98.29 & 0.5277 \\
    dualFFNN Entropy & 97.43 & 97.47 & 0.8819 & 99.06 & 98.94 & 0.5562 & \textbf{98.93} & 98.85 & 0.2571 \\
    dualFFNN k-medoids & \textbf{97.83} & \textbf{98.02} & 0.8230 & \textbf{99.11} & \textbf{99.05} & 0.3687 & 98.72 & 98.89 & 0.3980 \\
    dualFFNN MC Dropout & 97.67 & 97.90 & 0.8589 & 99.09 & 98.99 & 0.5917 & 98.89 & \textbf{98.90} & 0.2679 \\
    FFNN-TL Random & 90.63 & 89.44 & 0.9996 & 95.50 & 94.74 & 0.9968 & 97.07 & 96.66 & 0.9645 \\
    FFNN-TL Entropy  & 92.32 & 91.21 & 0.9995 & 97.24 & 96.50 & 0.9960 & 98.24 & 97.77 & 0.8079 \\
    FFNN-TL k-medoids  & 93.39 & 92.26 & 0.9986 & 97.33 & 96.58 & 0.9923 & 98.36 & 97.88 & 0.7123 \\
    FFNN-TL MC dropout & 89.98 & 88.78 & 0.9995 &  95.93 & 95.17 & 0.9972 & 97.62 & 97.20 & 0.9031 \\
    FFNN Random & 84.76 & 83.46 & 0.9998 & 95.23 & 94.48 & 0.9986 & 96.64 & 96.63 & 0.9793 \\
    FFNN Entropy  & 86.67 & 85.32 & 0.9994 &  97.31 & 96.56 & 0.9934 & 97.97 & 97.44 & 0.9260 \\
    FFNN k-medoids  & 90.13 & 88.74 & 0.9984 & 97.74 & 96.97 & 0.9942 & 98.17 & 97.64 & 0.7843 \\
    FFNN MC dropout & 90.20 & 88.83 & 0.9994 & 97.57 & 96.83 & 0.9830 & 98.11 & 89.36 & 0.7817 \\
    LGB Random & 90.76 & 89.25 & 0.9529 & 97.45 & 96.62 & 0.6981 & 97.45 & 96.97 & 0.4774 \\
    LGB Entropy  & 97.33 & 95.71 & 0.7867 & 98.66 & 98.08 & 0.3184 & 98.66 & 98.08 & 0.3184 \\
    LGB k-medoids  & 97.24 & 95.60 & 0.8323 & 98.92 & 98.11 & 0.4629 & 98.67 & 97.99 & 0.3095\\
    \midrule
    dualFFNN (200K) & \textbf{97.79} & \textbf{97.86} & 0.8812 & \textbf{99.00} & \textbf{99.03} & 0.5889 & \textbf{98.76} & \textbf{98.77} & 0.4484 \\
    FFNN-TL (200K) & 95.07 & 93.85 & 0.9991 & 97.66 & 96.93 & 0.9644 & 98.30 & 97.83 & 0.8261 \\
    FFNN (200K) & 92.15 & 90.77 & 0.9998 & 97.89 & 97.11 & 0.9988 & 98.11 & 97.56 & 0.8730 \\
    LGB (200K) & 95.85 & 94.19 & 0.9307 & 98.48 & 97.65 & 0.5868 & 98.32 & 97.71 & 0.3866 \\
    \bottomrule
    \end{tabular}
 \end{adjustbox}
\end{table*}

\subsection{Stealing the Sorel20m Models}
\label{sec:stealingsorel}
A subset of the Sorel20m dataset was used as the thief and test datasets against the two Sorel20m targets. All metrics were measured at an FPR equal to $0.002$ for the Sorel-LGB target and $0.006$ for the Sorel-FFNN target. These FPR levels were achieved using $0.5$ as a threshold for both targets. Each experiment was run five times to randomize the initial seed and validation data selection. The FPR levels for the Sorel models are significantly lower than the Ember model without having to adjust the decision threshold. This is probably because the models were created and tuned with a much larger dataset than Ember 2018. 

The results for the model extraction of the Sorel-LGB model are depicted in Figures~\ref{fig:sorel_LGB_agreement} and~\ref{fig:sorel_LGB_roc}. The dualFFNN and the LGB surrogates reach agreement scores of around 99\% with as few as 13,000 queries (0.0856\% of the samples used to train the target). The FFNN-TL and the vanilla FFNN are slightly lower in agreement, but they are worse when looking at the ROC curves. The FFNN-TL is better than the FFNN surrogate in the very low FPR regions. Table~\ref{tab:models_results} shows that the difference between the sampling strategies is small, and there is no clear winning non-random sampling strategy. It also shows that active learning strategies can create models that perform similarly to the target, which is trained with 20 million data points. 

For the Sorel-FFNN target, the agreement and ROC curves can be seen in Figures~\ref{fig:sorel_FFNN_agreement} and~\ref{fig:sorel_FFNN_roc} respectively. Again the LGB and dualFFNN surrogates achieve a high level of agreement, with the LGB surrogate having a ROC curve slightly closer to that of the target. 

An interesting observation for the attacks on both Sorel20m targets and the Ember2018 target is that the two Sorel20m targets performed slightly better than Ember2018 in the respective test sets. The Sorel20m models had an accuracy of $98.86$ and $98.68$ while Ember2018 had $96.50$. Moreover, the Sorel20m targets were slightly easier to steal than the Ember2018 target.

\subsection{Attacking Using a Different Data Distribution}
\label{sec:attacking-diff-distribution}
Sometimes, an attacker may have data from a different distribution than the one used to train the target or may not even know the original training data distribution. In the previous sections, we used thief and test datasets not previously seen by the targets, but we can assume they were roughly from similar distributions. 

To examine the effect of using different thief and test datasets, we created surrogates for the Sorel20m targets using the Ember thief and test datasets. Although they cover a similar chronological period, the two datasets overlap very little, with less than 14\% of the Ember dataset present in the much larger Sorel20m dataset. 

The Sorel20m target models were trained with millions of data points from the Sorel20m training dataset. However, they do not perform well on the Ember test set. To get some relatively low FPRs (0.08 for the Sorel-LGB and 0.11 for the Sorel-FFNN), we set the target model threshold to $0.9$. The results in Table~\ref{tab:sorel_results} indicate that it is possible to create LGB surrogates with up to 96\% agreement. However, the advantage of the dualFFNN architecture is no longer visible as it performs similarly to the FFNN architecture at around 95\% agreement. The extra dataset used for the FFNN-TL model does not seem to improve its performance, either. 

The experiment in this section showed that there is a performance penalty if the attacker does not possess data from the same distribution as the target model. Compared to the previous section, the surrogates did not reach as high agreement rates. However, the best surrogates reached 94\% agreement or higher. As expected, the attack is data-dependent. This behavior can be attributed to the fact that a model queried with data quite different from the training data distribution may behave unexpectedly, especially since the Ember feature space is so high dimensional. 

\begin{table*}[!t]
\footnotesize
    \caption{Metrics for Sorel-LGB and Sorel-FFNN surrogates trained using the Ember 2018 dataset at the 0.08 and 0.11 FPR level respectively with 25K query budget. The lower part are metrics for Sorel-LGB and Sorel-FFNN surrogates trained using the Ember 2018 dataset at the 0.08 and 0.11 FPR using 200K samples from the thief dataset without sampling.}
    \label{tab:sorel_results}
    \centering
    \begin{tabular}{l c c c | c c c }
    \toprule
    Target &  \multicolumn{3}{c|}{\textbf{Sorel-LGB}} & \multicolumn{3}{c}{\textbf{Sorel-FFNN}} \\
    Model \& Strategy & Agreement(\%) & Accuracy(\%) & Threshold & Agreement(\%) & Accuracy(\%) & Threshold  \\
    \midrule
    Target & - & 91.16 & 0.9000 & - & 90.45 & 0.9000 \\
    \midrule
    dualFFNN Random & 92.45 & 88.38 & 0.7684 & 93.72 & 89.63 & 0.6840  \\
    dualFFNN Entropy & 94.32 & 89.69 & 0.7504 & 95.55 & 90.65 & 0.3611 \\
    dualFFNN k-medoids & 94.86 & \textbf{90.01} & 0.6679 & 95.89 & 91.02 & 0.3356  \\
    dualFFNN MC Dropout & 94.68 & 89.97 & 0.5633 & 95.93 & \textbf{91.08} & 0.3541  \\
    FFNN-TL Random & 91.81 & 88.13 & 0.9716 & 92.60 & 88.61 & 0.9515  \\
    FFNN-TL Entropy & 93.22 & 89.38 & 0.9241 & 94.73 & 90.05 & 0.8034 \\
    FFNN-TL k-medoids  & 94.02 & 89.60 & 0.8163 & 94.96 & 90.04 & 0.7451  \\
    FFNN-TL MC dropout & 91.20 & 86.59 & 0.9884 & 92.10 & 87.56 & 0.9329 \\
    FFNN Random & 91.74 & 86.22 & 0.9791 & 91.92 & 87.13 & 0.9695 \\
    FFNN Entropy  & 93.62 & 87.66 & 0.9215 & 94.30 & 88.77 & 0.8701 \\
    FFNN k-medoids  & 94.67 & 88.73 & 0.8261 & 95.09 & 89.44 & 0.7975 \\
    FFNN MC dropout & 94.44 & 88.53 & 0.7842 & 94.85 & 89.36 & 0.7817 \\
    LGB Random & 93.96 & 87.64 & 0.6833 & 93.60 & 89.03 & 0.6276 \\
    LGB Entropy  & 96.28 & 89.19 & 0.4928 & 96.39 & 90.75 & 0.4364 \\
    LGB k-medoids  & \textbf{96.32} & 89.01 & 0.5137 & \textbf{96.45} & 90.74 & 0.4490 \\
    \midrule
    dualFFNN (200K) & \textbf{97.79} & \textbf{97.86} & 0.8812 & \textbf{96.11} & \textbf{90.86} & 0.4217 \\
    FFNN-TL (200K) & 94.44 & 89.79 & 0.8723 &  95.39 &90.11 & 0.7450  \\
    FFNN (200K) & 94.61 & 88.47 & 0.8376 & 95.29 & 89.43 & 0.7595  \\
    LGB (200K) & 95.77 & 88.51 & 0.6002 & 96.02 & 90.41 & 0.5153 \\
    \bottomrule
    \end{tabular}

\end{table*}

\subsection{Lessons Learned}
Model extraction attacks on the stand-alone models were successful, and the active learning strategies worked better than the random sample selection. Across all targets, we see that the entropy strategy performs similarly and sometimes better than more complex strategies (Table~\ref{tab:models_results}). This implies that the strategies do not need to be complicated or time-consuming. 

The sampling strategies work well in terms of the query budget used, and good surrogates, both in terms of agreement and accuracy, can be created with a small subset of the original training datasets. The dualFFNN reaches its peak agreement values with only 13,000 queries (Figure~\ref{fig:agreement_ember_sorel}).

The attack works well in terms of transferability, with both the dualFFNN and the LGB surrogates performing quite well against all types of target models. It is important to note that LightGBM models tend to perform quite well in sparse feature sets such as the Ember 2018. A high-performing neural network such as the dualFFNN trained with few data points is important because it allows more training flexibility than the LGB.

The most crucial factor for the attacks seems to be the selection of thief and test datasets and the performance of the targets on the data the attacker possesses. If a model performs well in the thief and test datasets, it will return labels closer to the true labels. In contrast, a low-performing target may return incorrect labels, making the training of the surrogates harder, similarly to training models with noisy data. 

\section{Stealing Antivirus Functionality}
\label{sec:stealing-av-functionality}
While it is undoubtedly interesting to test model stealing attacks on published models from a research perspective, these models do not necessarily reflect the state of machine learning used by various AV vendors. The motivation behind the experiments described in this section is to test whether it is possible to create surrogates of actual AVs that contain ML components and to which extent.

Performing model extraction on AVs is far more complicated than attacking published machine learning models. First, the attacker has no knowledge of malware preprocessing and the features used for malware detection. Second, even when a file scan is requested, the AV may perform several actions, such as unpacking or emulation, and there is little to no control over it. Third, the knowledge about the datasets used for training models is minimal. Finally, most AVs may contain multiple components used in file scans, such as pattern-based signatures, heuristics, etc. In most cases, there is no control over the use of these components, and the system must be viewed as a black box. In this work, we create surrogate models that learn the behavior of AV in a similar configuration to what an end-user would have.

To avoid bias in the analysis and for privacy reasons, the names of the AVs were anonymized, and only terms AV1 to AV4 were used.

\subsection{Experimental Setup}

Attacking AVs requires real files. Therefore we used our \textit{internal training} dataset as the thief dataset and the respective test set of the \textit{internal} dataset for the evaluation since the Ember 2018 dataset does not contain binaries. The surrogate models used were the same as described in Section~\ref{sec:stealing_stand_alone_ml_models}. The sampling strategies were also the same but used a query budget of $4,000$ and four query rounds. The smaller query budget is because the thief and test datasets were small, and we wanted to reduce the number of queries on the real AVs.

The number of queries required to train surrogate models is too high for a manual scan, so we automated the file scanning process by developing and installing a web service in each VM. The provided HTTP API allowed us to upload and scan multiple files automatically using the command line interface of each AV. 

We respected the default values for the AVs as much as possible, and we did not switch off any functionality related to file scanning. Finally, the VMs were disconnected from the internet to avoid updates during the experiments.

\subsection{Performance of AVs in the Internal Dataset}
Before performing any attacks, we measured the AV performance by scanning all the binary files of the internal dataset. Table~\ref{tab:AVs_results} shows the accuracy of each AV, as well as their TPR and FPR as of August 2022. All AVs had very few false positives in both training and test sets. Having low false positives is probably by design as they aim to reduce end-user inconvenience. 

\begin{table}[!t]
    \caption{Metrics for each AV using the internal dataset.}
    \label{tab:AVs_results}
    \centering
    \begin{tabular}{l c c c | c c c}
    \toprule
     & \multicolumn{3}{c |}{Thief dataset} &  \multicolumn{3}{c}{Test set} \\
     AV & Acc(\%) & FPR(\%) & TPR(\%) & Acc(\%) & FPR(\%) & TPR(\%)\\
    \midrule
    AV1 & 96.64 & 0.01 & 95.94 & 98.20 & 0 & 97.48 \\
    AV2 & 99.73 & 0.04 & 99.50 & 99.94 & 0.4 & 99.90 \\
    AV3  & 97.84 & 0.01 & 97.39 & 99.47 & 0.4 & 99.41 \\
    AV4 & 94.76 & 0 & 93.67 & 96.91 & 0.01 & 95.68\\
    \bottomrule
    \end{tabular}
\end{table}

The accuracy of three out of four AVs in the thief dataset was around 2\% lower than in the test set. This difference may be attributed to the time difference between the binaries in the two subsets. The test set contains binaries that were first seen after September 2019 and are more recent. Some of the AVs could decide not to "care" about malware as they become older and focus on newer threats. 

It should also be noted that this is not a test of the detection ability of each product. There are several design decisions that the product teams may have made that we are not aware of. For instance, some AVs may have responded differently had the files been executed instead of scanned. However, detecting malware files in the system is a desirable property, even when the system is not connected to the internet.

\subsection{Results}

\begin{figure*}[t!]
\centering
    \begin{subfigure}[b]{0.24\textwidth}
        \centering
        \includegraphics[width=\textwidth]{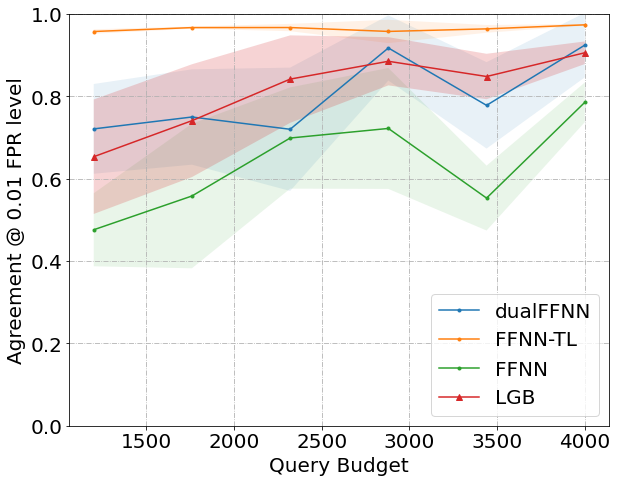}
        \caption{AV1}
        \label{fig:Avast_agreement}
    \end{subfigure}
     \hfil
    \begin{subfigure}[b]{0.24\textwidth}
        \centering
        \includegraphics[width=\textwidth]{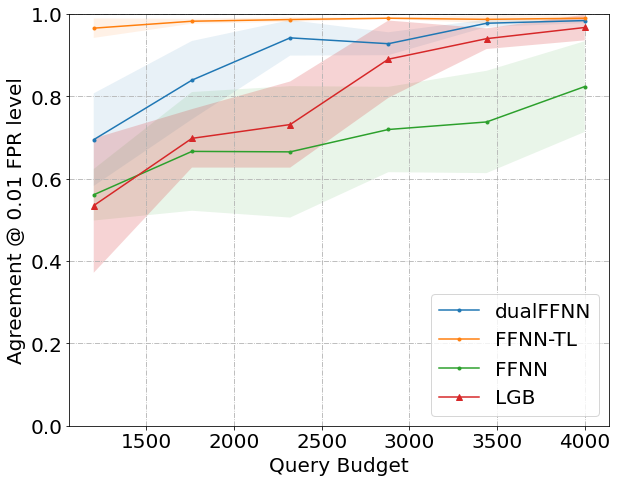}
        \caption{AV2}
        \label{fig:ESET_agreement}
    \end{subfigure}
    \hfil
    \begin{subfigure}[b]{0.24\textwidth}
        \centering
        \includegraphics[width=\textwidth]{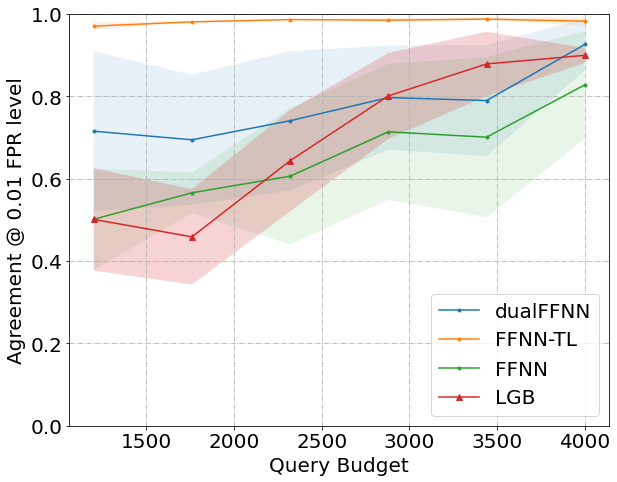}
        \caption{AV3}
        \label{fig:Kaspersky_agreement}
    \end{subfigure}
     \hfil
    \begin{subfigure}[b]{0.24\textwidth}
        \centering
        \includegraphics[width=\textwidth]{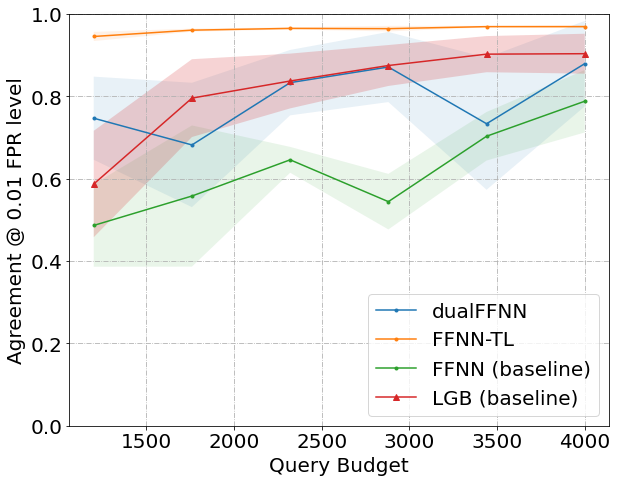}
        \caption{AV4}
        \label{fig:MS_agreement}
    \end{subfigure}
    \caption{Agreement of top 3 surrogate models+sampling strategy with the four AVs at 0.01 FPR.}
    \label{fig:agreement_all}
\end{figure*}

\begin{figure*}[t!]
\centering
    \begin{subfigure}[b]{0.24\textwidth}
        \centering
        \includegraphics[width=\textwidth]{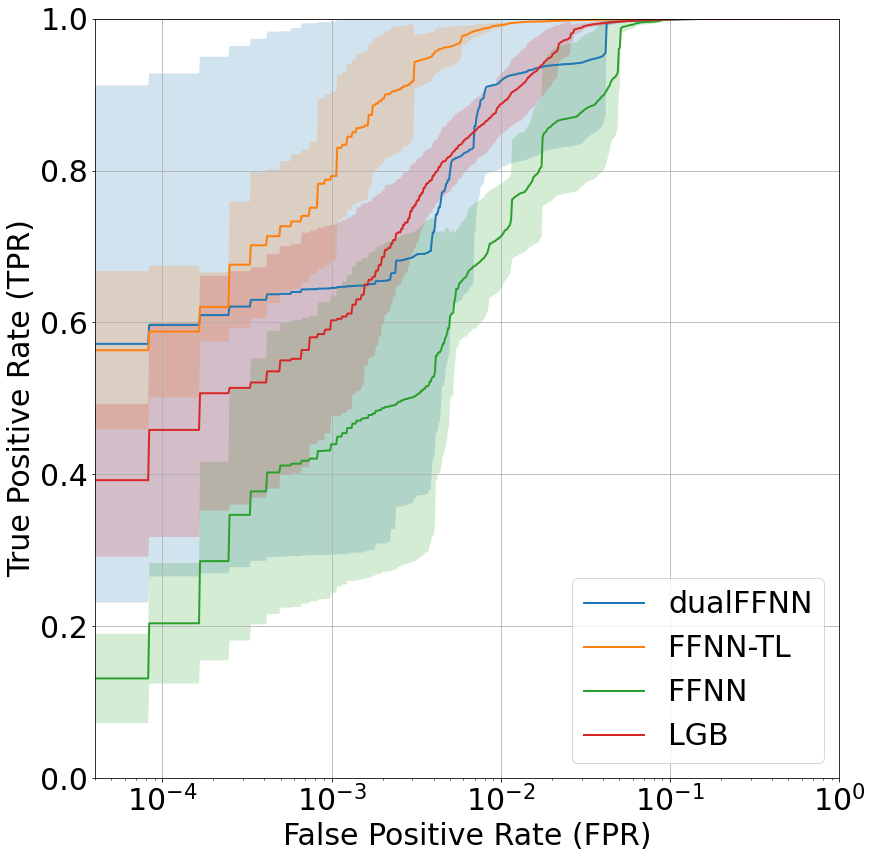}
        \caption{AV1}
        \label{fig:AV1_roc}
    \end{subfigure}
     \hfil
    \begin{subfigure}[b]{0.24\textwidth}
        \centering
        \includegraphics[width=\textwidth]{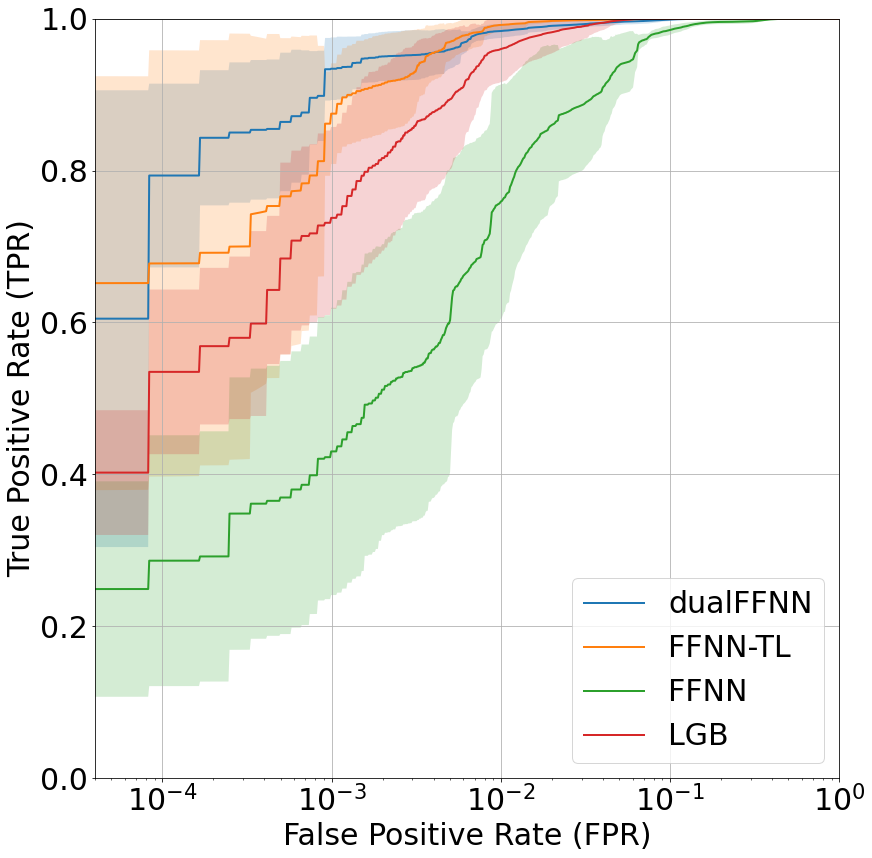}
        \caption{AV2}
        \label{fig:AV2_roc}
    \end{subfigure}
    \begin{subfigure}[b]{0.24\textwidth}
        \centering
        \includegraphics[width=\textwidth]{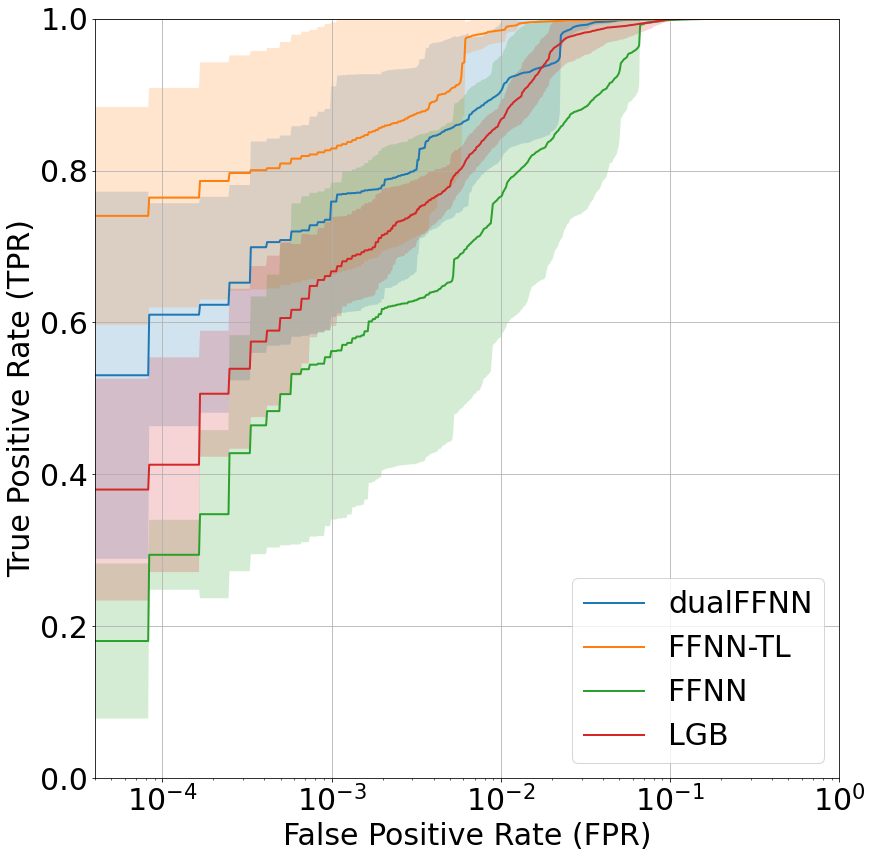}
        \caption{AV3}
        \label{fig:AV3_roc}
    \end{subfigure}
     \hfil
    \begin{subfigure}[b]{0.24\textwidth}
        \centering
        \includegraphics[width=\textwidth]{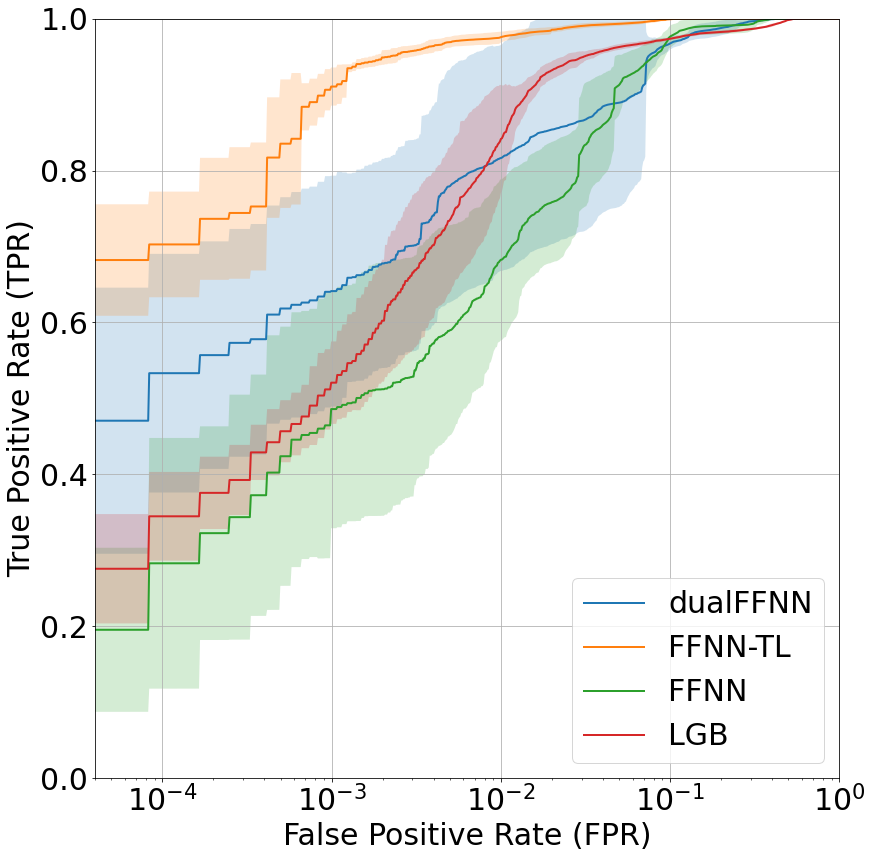}
        \caption{AV4}
        \label{fig:AV4_roc}
    \end{subfigure}
    \caption{ROC curves of top 3 surrogate models+sampling strategies for the four AVs.}
    \label{fig:roc_AVs_all}
\end{figure*}

\begin{table*}[!t]
    \caption{Metrics for the different surrogates attack models and strategies against all AVs at the 0.01 FPR level using 4K queries. The lower part are metrics for the surrogates using the full thief dataset (no sampling) at 0.01 FPR level using 140K query budget.}
    \label{tab:AV_results}
    \centering
    \begin{adjustbox}{width=\textwidth}
    \begin{tabular}{l c c c | c c c | c c c | c c c }
    \toprule
    Target / &  \multicolumn{3}{c|}{\textbf{AV1}} & \multicolumn{3}{c|}{\textbf{AV2}} & \multicolumn{3}{c}{\textbf{AV3}} & \multicolumn{3}{c}{\textbf{AV4}} \\
    Model \& Strategy & Agrmt.(\%) & Acc.(\%) & Thresh. & Agrmt.(\%) & Acc.(\%) & Thresh. & Agrmt.(\%) & Acc.(\%) & Thresh. & Agrmt.(\%) & Acc.(\%) & Thresh. \\
    \midrule
    Target & - & 98.20 & - & - & 99.94 & - & - & 99.47 & - & - & 96.91 & - \\
    \midrule
    dualFFNN Random & 73.10 & 73.17 & 0.9654 & 92.40 & 92.53 & 0.9847 & 83.47 & 83.59 & 0.9744 & 75.81 & 74.12 & 0.9514 \\
    dualFFNN Entropy & 92.44 & 93.85 & 0.8141 & 87.63 & 87.75 & 0.9629 & 89.26 & 89.57 & 0.9541 & 80.80 & 79.46 & 0.9189 \\
    dualFFNN k-medoids  & 84.63 & 85.50 & 0.8704 & 98.40 & 98.55 & 0.9644 & 91.63 & 91.92 & 0.9340 & 87.24 & 85.76 & 0.9391 \\
    dualFFNN MC dropout & 81.67 & 82.48 & 0.9552 & 97.08 & 97.22 & 0.9539 & 92.60 & 92.91 & 0.9141 & 87.95 & 86.59 & 0.8873 \\
    FFNN-TL Random & 96.82 & 98.49 & 0.3204 & 98.34 & 98.48 & 0.3961 & 98.12 & 98.43 & 0.4038 & 95.43 & 95.19 & 0.5581 \\
    FFNN-TL Entropy & 96.58 & 98.22 & 0.6069 & \textbf{99.00} & \textbf{99.15} & 0.5144 & \textbf{98.58} & \textbf{98.94} & 0.6084 & 96.21 & 96.04 & 0.5953 \\
    FFNN-TL k-medoids  & \textbf{97.37} & \textbf{99.11} & 0.4805 & 98.97 & 99.12 & 0.4140 & 98.25 & 98.62 & 0.4736 & \textbf{96.93} & \textbf{97.98} & 0.3135 \\
    FFNN-TL MC dropout & 96.51 & 98.15 & 0.4705 & 98.66 & 98.81 & 0.4031 & 97.89 & 98.24 & 0.4660 & 94.03 & 94.57 & 0.5120 \\
    
    FFNN Random & 63.04 & 62.42 & 0.9899 & 82.38 & 82.50 & 0.9969 & 70.36 & 70.27 & 0.9926 & 55.58 & 53.19 & 0.9880 \\
    FFNN Entropy & 66.72 & 66.52 & 0.9772 & 76.88 & 76.99 & 0.9985 & 82.79 & 82.96 & 0.9743 & 65.95 & 63.75 & 0.9886 \\
    FFNN k-medoids  & 78.02 & 78.47 & 0.9538 & 79.78 & 79.90 & 0.9972 & 74.22 & 74.26 & 0.9862 & 65.03 & 63.14 & 0.9895 \\
    FFNN MC dropout & 78.55 & 79.22 & 0.9613 & 79.36 & 79.48 & 0.9942 & 72.20 & 72.23 & 0.9828 & 78.82 & 77.01 & 0.9647 \\
    LGB Random & 79.33 & 79.43 & 0.9586 & 80.02 & 80.13 & 0.9979 & 78.81 & 78.85 & 0.9747 & 81.32 & 79.23 & 0.9548 \\
    LGB Entropy  & 88.39 & 89.46 & 0.9241 & 96.72 & 96.85 &  0.9884 & 89.96 & 90.21 & 0.9505 & 73.64 & 71.26 & 0.9668 \\
    LGB k-medoids  & 90.59 & 91.69 & 0.9315 & 95.38 & 95.52 & 0.9962 & 88.94 & 89.18 & 0.9601 & 90.36 & 88.39 & 0.9238 \\
    \midrule 
    dualFFNN (140K) & 95.69 & 96.97 & 0.8180 & 98.49 & 98.63 & 0.9870 & 89.57 & 89.76 & 0.9280 & 88.21 & 86.16 & 0.8340 \\
    FFNN-TL (140K) & \textbf{96.67} & \textbf{98.19} & 0.7444 & \textbf{99.16} & \textbf{99.31} & 0.7651 & \textbf{98.55} & \textbf{98.90} & 0.7424 & \textbf{97.16} & \textbf{95.81} & 0.7318 \\
    FFNN (140K) & 69.28 & 68.72 & 0.9598 & 74.82 & 74.92 & 0.9995 & 83.67 & 83.55 & 0.9649 & 73.64 & 71.26 & 0.9668 \\
    LGB (140K) & 87.53 & 88.15 & 0.9302  & 97.93 & 98.07 & 0.9935 & 88.33 & 88.32 & 0.9540 & 89.05 & 86.78 & 0.9304 \\
    \bottomrule
    \end{tabular}
    \end{adjustbox}
\end{table*}

The results of the experiments using AVs 1-4 as black-box targets are presented in Table~\ref{tab:AV_results} and Figures~\ref{fig:agreement_all} and~\ref{fig:roc_AVs_all}. The FFNN-TL surrogates perform much better than the other surrogates in terms of agreement for all AVs. In addition, the method shows a lot less variance in the results compared to dualFFNN and the baseline models. The high variance of the other surrogate models can be attributed to the small and imbalanced dataset, which could be a real problem for an attacker. This is also apparent in the ROC curves in Figure~\ref{fig:roc_AVs_all}. The best results in agreement and surrogate model accuracy are against AV2, which was the AV with the smallest gap in performance between the thief and test datasets (Table~\ref{tab:AVs_results}). When using the full dataset (140K data points and no sampling strategy), the agreement of the FFNN-TL surrogate is higher for all AVs, and the dualFFNN performs quite well for both AV1 and AV2. The advantage of having additional data, even if not used to query the target, is quite significant.

\subsection{Lessons Learned}
AVs are more challenging to steal than stand-alone models, probably because (i) they have been trained with unknown and more extensive datasets and (ii) because they are more complex as detection models, as was apparent in their log files. It is significantly more challenging because an AV may employ signatures and heuristics that are not easily mapped in the feature set used. However, getting agreement scores of 97\% or more is feasible if the attacker uses additional data, even if they are not in the form of actual binary files and cannot be used to query the AVs. Additional data offer more stability in the training process, and in general, the FFNN-TL model reached high agreement scores with relatively fewer queries.

\section{Creating Adversarial Malware}
\label{sec:creating-adversarial-malware}

\subsection{Experimental Setup}
After creating the surrogates for AVs, we tested their efficacy for generating adversarial malware binaries using two different attacks: MAB and GAMMA. The MAB attack~\cite{song2021mabmalware} is based on reinforncement learning and it's goal is to select the most appropriate modification (action) to a binary file based on the output of the detection model or AV. The attack was used in its original settings, with the only difference being that we did not use the code randomization action since it required access to proprietary software. MAB requires benign PE sections, and for that purpose, we extracted $20,000$ sections from the benign binaries in our internal dataset. 

GAMMA is using genetic algorithms to select benign sections that are appended to the malware binary in order to make it look less malicious. For the GAMMA attack, we used only $30$ sections because the algorithm does not scale well as the number of benign sections increases. The names of the sections were randomly generated. The number of iterations for GAMMA was  $10$, and the population size was $20$, which is 210 queries for each candidate solution. The $\lambda$ of the attack that affects the size of the injected data was $10^{-6}$. 

A thousand malicious binaries were randomly chosen from a subset overlapping the Ember test set and our internal dataset for the MAB attacks. Although the data selection is biased in favor of the Ember model and its surrogates, we opted to use malicious binaries that belonged to families that were also part of the internal dataset. All experiments were performed using these same malware files, and each experiment was repeated three times to account for the randomness of the modifications performed by MAB. We only repeated the experiment once for the AVs since the attack was much slower than on ML models. In addition, for AV4, the attack did not conclude after 24 hours, and we decided to stop it and report the number of malware binaries that had become evasive until then. 

The GAMMA attack is much slower than the MAB attack. Therefore we reduced the number of malware binaries to $200$ per seed. These binaries were a subset of the thousand binaries used in the MAB attack.

\subsection{Surrogates vs. Targets}
\begin{figure*}[t!]
\centering
    \begin{subfigure}[b]{0.45\textwidth}
        \centering
        \includegraphics[width=\textwidth]{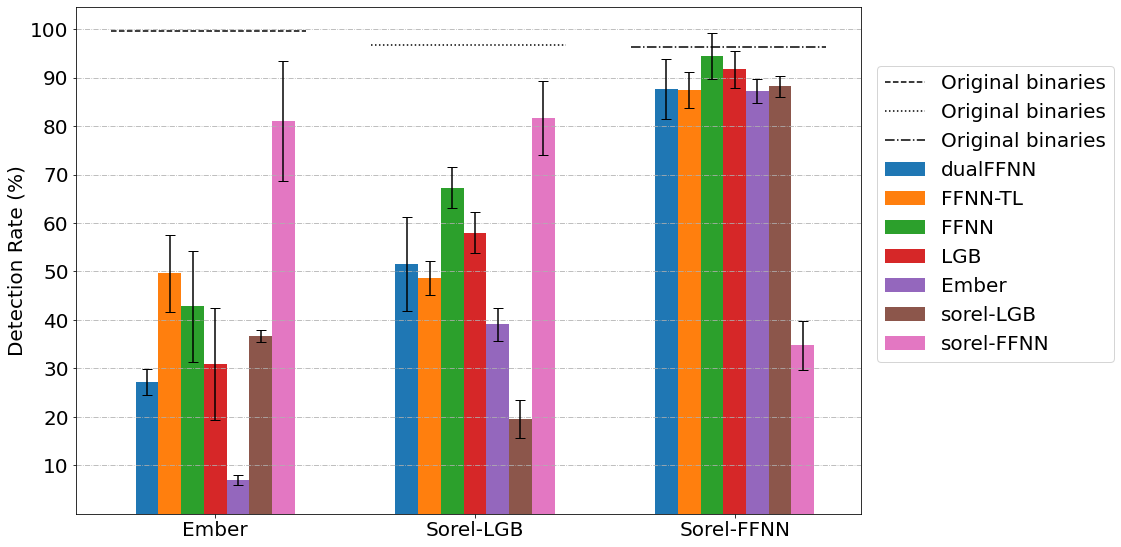}
        \caption{Stand-alone Models}
        \label{fig:MAB_models}
    \end{subfigure}
     \hfil
    \begin{subfigure}[b]{0.45\textwidth}
        \centering
        \includegraphics[width=\textwidth]{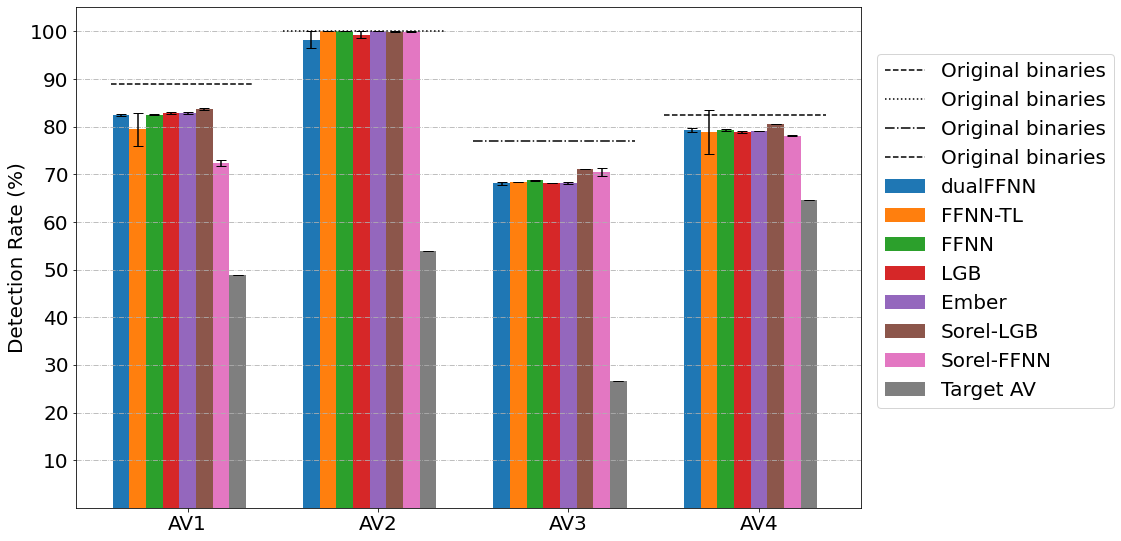}
        \caption{AVs}
        \label{fig:MAB_AVs_minimal}
    \end{subfigure}
    \caption{Detection rates of the adversarial malware generated by MAB, grouped by target. The horizontal lines at the top are the detection rates for the unmodified original malware. The three leftmost bars in each group represent the detection rates of the best surrogates of each target. The three rightmost bars in each group are the detection rates of the three target models.}
    \label{fig:MAB_all}
\end{figure*}

\begin{figure*}[t!]
\centering
    \begin{subfigure}[b]{0.45\textwidth}
        \centering
        \includegraphics[width=\textwidth]{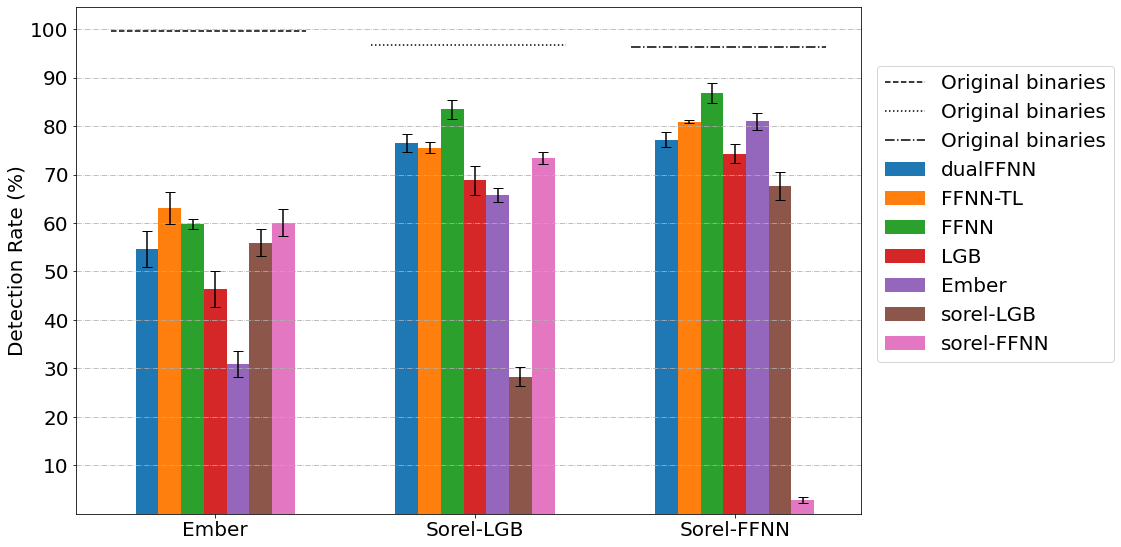}
        \caption{Stand-alone Models}
        \label{fig:GAMMA_models}
    \end{subfigure}
     \hfil
    \begin{subfigure}[b]{0.45\textwidth}
        \centering
        \includegraphics[width=\textwidth]{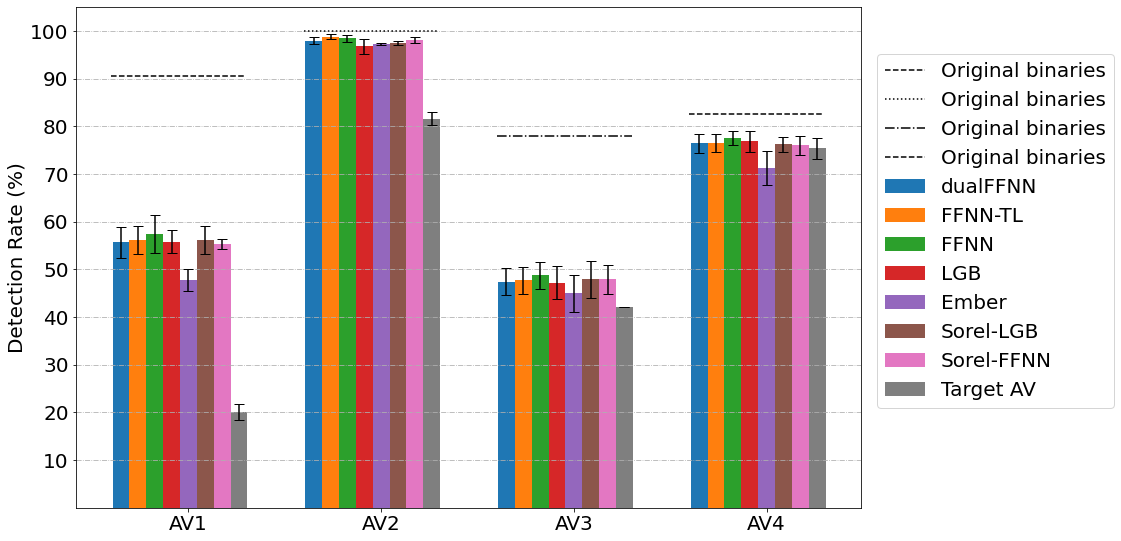}
        \caption{AVs}
        \label{fig:GAMMA_AVs_minimal}
    \end{subfigure}
    \caption{Detection rates of the adversarial malware generated by GAMMA, grouped by target. The horizontal lines at the top are the detection rates for the unmodified original malware. The three leftmost bars in each group represent the detection rates of the best surrogates of each target. The three rightmost bars in each group are the detection rates of the three target models.}
    \label{fig:GAMMA_all}
\end{figure*}

Using the setup described above, we generated adversarial malware samples using the best surrogates based on agreement scores, the stand-alone models, and the AVs. For a better comparison we also used the target AV models to create adversarial malware. We computed the Ember features for each binary sample to test the stand-alone models. The metric used to evaluate the results was the mean detection rate of each target on the adversarial samples. The results are shown in Figures~\ref{fig:MAB_all} for the MAB algorithm and in Figure~\ref{fig:GAMMA_all} for the GAMMA algorithm. Each target is shown on the $x$ axis. The four leftmost bars for each target (blue, orange, green, red) depict the detection rates of the best surrogates for each respective target. The next three leftmost bars for each target show the detection rates when using the same targets to create adversarial samples against themselves. As a baseline reference, the figure also shows a top dotted line for the detection rates of the \textit{unmodified and not adversarial} original malware samples.

In the attacks against the stand-alone models (Figures~\ref{fig:MAB_models} and~\ref{fig:GAMMA_models}), we can observe that the dualFFNN surrogate achieved the lowest detection rates compared to the other surrogates in most cases for the MAB attack. In contrast, the LGB surrogate was best in the GAMMA attack. The attacks against the AVs (Figures~\ref{fig:MAB_AVs_minimal} and~\ref{fig:GAMMA_AVs_minimal}) had very similar results for all surrogates. AV2 and AV4 were the most robust against adversarial samples generated by both attack frameworks. AV1 and AV3 were much more vulnerable to the GAMMA attack than to the MAB one. 

It is also evident that generating evasive malware using surrogates is easier for the Ember and Sorel-LGB targets than for some AVs and the Sorel-FFNN targets. While the results concerning the AV targets may be explained by the fact that creating surrogates for such complex black-boxes is a challenging task, the Sorel-FFNN results indicate that for these two attacks, it may be harder to evade a neural network target. This can be attributed to the potentially more complex decision boundary of neural networks that might not be easily learned by the surrogates. 

Figures ~\ref{fig:MAB_AVs_minimal} and~\ref{fig:GAMMA_AVs_minimal} shows that the target AV were better than their surrogates for creating adversarial malware against themselves. Even though this is expected, it should be considered when surrogates are required as an intermediate step for another attack. An argument favoring surrogates despite their lower performance is that using an AV to create adversarial malware can be too costly. In addition, the two attack frameworks we used did not take advantage of the fact that some of the surrogates can be used as white-boxes and potentially generate better adversarial malware. Attacks that can use gradient-based information may be able to take advantage of the full potential of the neural network surrogates. 

Both MAB and GAMMA attacks use black-box targets, but when used against the AVs, they were extremely time-consuming. In particular, the MAB attack against AV4 had to be stopped after 24 hours of generating malware. Attacking an AV directly might be less attractive to an attacker if they need a faster and stealthy attack. AV vendors may also decide to slow down the attackers by introducing delays in their response times. These delays would make model stealing attacks with surrogates more attractive to attackers.

One question that may be relevant is whether using surrogates for adversarial malware generation is more effective than using existing public stand-alone models such as the Ember and Sorel20m models. To answer this question, we included in Figures~\ref{fig:MAB_all} and~\ref{fig:GAMMA_all} the detection rates of the adversarial malware created by the three stand-alone models for all the targets (three right-most columns in each group). Not counting each target model against itself, we see that the stand-alone models have similar performance with the rest of the surrogates, especially regarding the AV targets, with only the Ember2018 model being slightly more effective against AV1 and AV4 for the GAMMA attack. However, given that the adversarial malware subset is chosen from malware samples that are interesting to the attacker, as time passes, the stand-alone (fixed trained) models will not be able to adapt and detect newer malware, which suggests that surrogates are a better choice.

Finally, as mentioned in Section~\ref{sec:methodology}, the MAB framework works in two stages. In the first stage, it produces \textit{evasive} malware, and in the second stage, it reduces the number of modification actions to produce \textit{minimal} but still evasive malware. In Figure~\ref{fig:MAB_all} we show the detection results of the final minimal malware binaries. When we compared we saw that the minimal binaries were less detected by the stand-alone models. However, the evasive ones were less detected by the AVs. This suggests that learning the decision boundary of the AVs is a more challenging task, and as the binaries are modified less, they may evade the surrogates but not the targets themselves.

\subsection{Offline vs. Online Detection}
\label{sec:online_vs_offline}

\begin{table*}[!tp]
\footnotesize
\centering
\caption{Online vs. offline AV adversarial detections using MAB attack. Mean detection rates of all AVs against the original malware samples and the adversarial malware samples created by different attack models. "Orig" denotes the detection rate of the original unmodified malware. Offline means not connected to the Internet, while online means after connection to the Internet.}
\label{tab:AVs_online_MAB}
\begin{adjustbox}{width=0.45\textwidth}
\begin{tabular}{ccccccc}
\toprule
 Target & Attacker & Original & dualFFNN & FFNN-TL &  FFNN & LGB  \\ 
\midrule
\multirow{2}{3em}{AV1}   & Offline & 89\% & 82.4\% & 79.46\% & 82.5\% & 82.8\% \\
   & Online & 88.9\% & 82.2\% & 80.6\%& 82.3\% & 82.5\% \\
\midrule
\multirow{2}{3em}{AV2}   & Offline & 100\% & 98.3\% &100\% &  100\% & 99.3\% \\
   & Online & 100\%& 99.6\%& 99.97\%& 100\%& 99.8\% \\
\midrule
\multirow{2}{3em}{AV3}   & Offline & 77.1\% & 68.1\% & 68.37\%&  68.8\% & 68.2\% \\
   & Online & 99.3\%& 95.8\%& 95.63\% &  95.5\%&95.8\% \\
\midrule
\multirow{2}{3em}{AV4}   & Offline & 82.5\% & 79.3\% & 78.87\% & 79.4\% & 78.8\% \\
   & Online & 99.9\% & 96.3\%& 94.2\% & 96.6\%& 99.8\% \\
\bottomrule
\end{tabular}
\end{adjustbox}
\end{table*}

\begin{table*}[!tp]
\footnotesize
\centering
\caption{Online vs. offline AV adversarial detections using GAMMA attack. Mean detection rates of all AVs against the original malware samples and the adversarial malware samples created by different attacker models. "Orig" denotes the detection rate of the original unmodified malware. Offline means not connected to the Internet, while online means after connection to the Internet.}
\label{tab:AVs_online_GAMMA}
\begin{adjustbox}{width=0.45\textwidth}
\begin{tabular}{ccccccc}
\toprule
 Target & Attacker & Original & dualFFNN & FFNN-TL &  FFNN & LGB  \\ 
\midrule
\multirow{2}{3em}{AV1}   & Offline & 89\% & 55.64\% & 56.15\% & 57.5\% & 55.81\% \\
   & Online & 88.9\% & 58.17\% & 58.18\%& 58.35\% & 59.70\%\\
\midrule
\multirow{2}{3em}{AV2}   & Offline & 100\% & 97.98\% & 98.82\%&  98.48\% & 96.79\% \\
   & Online & 100\%& 97.97\%& 97.81\%& 98.99\%& 97.97\% \\
\midrule
\multirow{2}{3em}{AV3}   & Offline & 77.1\% & 47.39\% & 47.72\% & 48.74\% & 47.22\% \\
   & Online & 99.3\%& 82.3\%& 82.12\%&  83.48\%& 83.14\% \\
\midrule
\multirow{2}{3em}{AV4}   & Offline & 82.5\% & 76.4\% & 76.56\%& 77.57\% & 76.9\% \\
   & Online & 99.9\% & 99.9\%& 99.9\%& 99.66\%& 99.9\%  \\
\bottomrule
\end{tabular}
\end{adjustbox}
\end{table*}

One of the limitations of studying actual AVs is that experiments need to be stable and repeatable, and therefore we can not (at first) allow AVs to connect to the internet and update their databases and models. Therefore our experiments were conducted first with AVs offline and then connecting them once and trying the detection of adversarial malware again.

To provide a complete view of the efficacy of our adversarial samples, we tested the original and adversarial malware binaries after connecting the AVs to the internet and allowing them to update their models and databases. For the MAB attack, each AV was retested using 13,000 binaries in total: 3,000 adversarially generated with each of the four surrogates of that AV, and the 1,000 original binaries, as presented in Figure~\ref{fig:MAB_all}. For the GAMMA attack, each AV was tested with 600 binaries per surrogate and 600 original files for a total of 3,000 malware binaries.

Table~\ref{tab:AVs_online_MAB} shows the mean detection rates for the online and offline tests for MAB. Three out of four AVs manage to improve significantly when connected to the internet and using their cloud capabilities. Only AV1 had similar results in both offline and online settings. However, we must note that some of the AVs required significant time to scan all the binaries. Similarly, Table~\ref{tab:AVs_online_GAMMA} shows the results for the adversarial malware generated using the GAMMA attack. The results show very similar trends with the MAB attack. However, we can also notice that the detection levels of AV1 and AV3 are lower than the rest even \textit{after} they connect to the internet. This may indicate that both AVs are more susceptible to the GAMMA attack, which increases the number of sections in the binary.

\subsection{Lessons Learned}
Better surrogates in the agreement task tend to perform better in the adversarial malware creation task. The performance difference was more prominent when the targets were stand-alone models. Even though the attack results showed relatively high variance, dualFFNN, and FFNN-TL surrogates performed better than the FFNN surrogates in all cases except the Ember target in both MAB and GAMMA attacks.

In addition, the overall results show that high agreement and good ROC curves alone are not metrics that can guarantee the success of adversarial attacks. This is most evident for the Sorel-FFNN target in MAB attack, where even the high agreement surrogates did not generate many adversarial samples that can evade their target. 

Using existing models such as Ember or Sorel20m models instead of surrogates is not a strategy that is viable for an actual attacker, especially when going against real-world targets such the AVs. Using the AVs directly is not necessarily the preferred choice if the attack has time constraints, due to the fact that both attacks take a lot more time to complete.

Finally, the AVs tested can detect more than 68\% of the adversarial samples created by the MAB framework and two out of four AVs perform well against the GAMMA attack. When connected to the internet, three out of four AVs use cloud capabilities and increase their detection rates significantly.

\section{Discussion}
\label{sec:discussion}

Most model stealing attacks, as well as a lot of adversarial attacks, assume access to soft labels. However, malware classifiers used in actual products provide only hard labels. Our experiments show that it is possible to create surrogate models from hard-labeled black-box models and AVs, even at low FPR settings. However, surrogates have a drop in performance for the creation of adversarial malware as it was observed using two of the most powerful black-box attacks in the literature.

Surrogate model creation attacks proposed in the security domain literature do not extensively research the datasets used, ending up using parts of the same dataset for the surrogate and the target training. Even though our results show that a surrogate trained on a dataset similar to the one used by the target is more successful, we believe that the assumption that an attacker has the same dataset as its target is unrealistic. In Subsection~\ref{sec:attacking-diff-distribution} we measured datasets with little overlap, and the results showed that surrogates created with very different data reached lower levels of agreement between surrogates and targets.  

High agreement and high accuracy on a test set can translate to a better performance in the adversarial attacks, with our dualFFNN and FFNN-TL models outperforming the FFNN model in most targets. However, factors such as the type of target model can be significant. This was seen in the adversarial attacks on the two Sorel models, where Sorel-LGB is easier to attack under the same conditions than Sorel-FFNN. A future research avenue can be how to close this gap, requiring better metrics and datasets. Improved model stealing methods that use strategies that create valid synthetic binaries may also be an exciting approach. 

Regarding the adversarial malware generation without surrogates, results showed that stand-alone and AV models generated more evasive malware than their surrogates. However, the creation of adversarial malware using the AVs directly was very slow, and most AVs quickly detected the adversarial malware in the online setting. There are good reasons for using surrogates as a first step instead of directly attacking the targets:  a) most modern AVs use cloud-based detection as witnessed in Section~\ref{sec:online_vs_offline} and creating surrogates may provide a better estimation of detection ability while being stealthy; b) a surrogate can test thousands of files very fast, which is desirable to many "Malware as a service" providers that already exist~\cite{cybercrime2021sembera}, and c) the surrogate approach allows the attackers to test multiple AVs, which is also a possible use case for automating malware checks by service providers. Future work on adversarial attacks should consider time constraints and stealthiness. 

Finally, both of the adversarial malware generation methods we tested use models as a black or gray box and they don't take advantage of the surrogates as white-box models.
In addition, the currently known white-box attacks are primarily applicable to attacks against the MalConv architecture or similar neural network that take byte-embeddings as input features~\cite{demetrio2020adversarial}. Creating attacks that can use white-box models and manage to map feature-based malware modifications to the problem space has focused mainly on Android malware~\cite{intriguing2020}. We believe that being able to use the surrogates as white-box models for creating adversarial Windows malware would be an interesting future research direction.

\section{Conclusion}
\label{sec:conclusion}

This research explored training multiple types of surrogate models and sampling strategies to steal stand-alone machine learning models and four antivirus systems. The best surrogates achieved up to 99\% agreement using less than 25,000 training samples against stand-alone models trained with millions of data points. Our proposed dualFFNN architecture had not only top agreement scores, but it also performed well under very low FPR settings, closely followed by the LGB surrogates. Our FFNN-TL architecture achieved 96\%-99\% agreement only using 4,000 samples against antivirus products. 

The surrogates were used to create adversarial malware using the MAB and GAMMA attack frameworks. Regarding agreement scores, better surrogates generated lower detection rates in general. However, some targets were hard to evade, and all surrogates had similar performances. Evasion was easier against stand-alone LGB models (Ember and Sorel) than against Sorel-FFNN and the AVs for the MAB attack. The GAMMA attack was also quite successful against two of the AVs. Both attacks were most successful when the targets were used to generate adversarial malware directly. However, in the case of AVs, they were not time efficient enough to be practical, which is an advantage of surrogate models. Furthermore, neither of the two attacks (MAB and GAMMA) is white-box, and they cannot take advantage of the fact that we can obtain high-performing neural network surrogates. Creating better white-box attacks in the malware detection domain is an interesting future line of work.

The discrepancy between offline and online detection of the AVs was quite significant, with some AVs reporting less than 50\% detection rate, especially for the GAMMA attack. Three out of four AVs got better detection rates of adversarial malware after connecting to the internet, showing that cloud-based detection is an essential aspect of today's AVs. This further strengthens the need for surrogate creation if the attackers are interested in being more stealthy in their operations. 

\section{Acknowledgements}
This work was partially supported by Avast Software and the OP RDE funded project Research Center for Informatics No.: CZ.02.1.01/0.0./0.0./16\_019/0000765. The authors gratefully acknowledge the support of NVIDIA Corporation with the donation of a Titan V GPU used for this research.

\bibliographystyle{elsarticle-num}
\bibliography{stealing_models}

\end{document}